\documentclass[10pt,conference]{IEEEtran}

\usepackage{arydshln}
\usepackage[font=small,labelfont=bf,tableposition=top]{caption}
\usepackage{hyperref}
\usepackage{boldline}
\usepackage{color,colortbl}
\usepackage{bigstrut}
\usepackage[ruled]{algorithm2e}
\usepackage{amsmath,amsfonts}
\usepackage{amsmath}
\usepackage{array}
\usepackage{algorithmic}
\usepackage{balance}
\usepackage{blindtext}
\usepackage{booktabs}
\usepackage{caption}
\usepackage{hyperref}
\usepackage[noabbrev]{cleveref}
\usepackage{cite}
\usepackage{comment}
\usepackage{enumitem}
\usepackage{eqparbox}
\usepackage{fancybox}
\usepackage{fancyvrb}
\usepackage{framed}
\usepackage{graphicx}
\usepackage{ifthen}
\usepackage{listings}
\usepackage{mathrsfs}
\usepackage{mdwmath}
\usepackage{mdwtab}
\usepackage{multirow}
\usepackage{pifont}
\usepackage{stfloats}
\usepackage{textcomp}
\usepackage{times}
\usepackage{url}
\usepackage{xspace}
\usepackage[normalem]{ulem}
\usepackage[framemethod=TikZ]{mdframed}
\usepackage{caption}
\usepackage{subcaption}
\usepackage{tabularx}
\usepackage[switch]{lineno}

\usepackage{titlesec}

\usepackage{tcolorbox}
\tcbuselibrary{listings,skins}
 
\usepackage{mathtools}
\usepackage{blindtext}
\usepackage{lstautogobble}

\DeclareGraphicsExtensions{.pdf,.jpeg,.png}
\graphicspath{{images/}}

\titlespacing\section{0pt}{8pt plus 4pt minus 2pt}{4pt plus 2pt minus 2pt}

\newcommand{\rom}[1]{\uppercase\expandafter{\romannumeral #1\relax}}

\newcommand{\etal}{\hbox{\emph{et al.}}\xspace}
\newcommand{\eg}{\hbox{\emph{e.g.,}}\xspace}
\newcommand{\ie}{\hbox{\emph{i.e.,}}\xspace}

\newcommand{\wrt}{\hbox{\emph{w.r.t.}}\xspace}

\definecolor{gray50}{gray}{.5}
\definecolor{gray40}{gray}{.6}
\definecolor{gray30}{gray}{.7}
\definecolor{gray20}{gray}{.8}
\definecolor{gray10}{gray}{.9}
\definecolor{gray05}{gray}{.95}
\definecolor{gray01}{gray}{.97}

\newlength\Linewidth
\def\findlength{\setlength\Linewidth\linewidth
\addtolength\Linewidth{-4\fboxrule}
\addtolength\Linewidth{-3\fboxsep}
}
\newenvironment{examplebox}{\par\begingroup
  \setlength{\fboxsep}{3pt}\findlength
  \setbox0=\vbox\bgroup\noindent
  \hsize=0.90\linewidth
  \begin{minipage}{0.90\linewidth}\normalsize}
    {\end{minipage}\egroup
    \textcolor{gray20}{\fboxsep1.2pt\fbox
     {\fboxsep5pt\colorbox{gray05}{\normalcolor\box0}}}
    \endgroup\par\noindent
    \normalcolor\ignorespacesafterend}

\usetikzlibrary{shadows}
\usepackage{graphics}
\newmdenv[
    tikzsetting= {fill=blueish},
    skipabove=0.33em,
    skipbelow=0.33em,
    linewidth=1pt,
    innerleftmargin=4pt,
    innerrightmargin=4pt,
    innertopmargin=2pt,
    innerbottommargin=2pt,
    linecolor=gray95,
    roundcorner=2pt, 
    shadow=true,
    shadowsize=4pt,
    shadowcolor=gray95
]{questionbox}

\newmdenv[
    tikzsetting= {fill=greenish},
    skipabove=0.33em,
    skipbelow=0.33em,
    linewidth=1pt,
    innerleftmargin=4pt,
    innerrightmargin=4pt,
    innertopmargin=2pt,
    innerbottommargin=2pt,
    linecolor=gray95,
    roundcorner=2pt, 
    shadow=true,
    shadowsize=4pt,
    shadowcolor=gray95
]{answerbox}

\newmdenv[
    skipabove=0.33em,
    skipbelow=0.33em,
    innerleftmargin=4pt,
    innerrightmargin=4pt,
    innertopmargin=2pt,
    innerbottommargin=2pt,
]{lessonbox}

\usepackage{tikz}

\newenvironment{lesson}
{
    \begin{lessonbox}
}
{\end{lessonbox}}

\newenvironment{result}
{\begin{answerbox}}
{\end{answerbox}}

\newenvironment{question}
{\begin{questionbox}}
{\end{questionbox}}

\definecolor{javared}{rgb}{0.6,0,0} 
\definecolor{javagreen}{rgb}{0.25,0.5,0.35} 
\definecolor{javapurple}{rgb}{0.5,0,0.35} 
\definecolor{javadocblue}{rgb}{0.25,0.35,0.75} 

\lstdefinestyle{basejava}{
  language=java,
  showstringspaces=false,
  basicstyle=\scriptsize\ttfamily,
  keywordstyle=\bfseries\color{javapurple},
  commentstyle=\itshape\blue,
  identifierstyle=\blue,
  frame=none,
  backgroundcolor=\color{white},
}

\lstdefinestyle{CustomJava}{
  belowcaptionskip=\baselineskip,
  breaklines=true,
  xleftmargin=\parindent,
  language=java,
  showstringspaces=false,
  basicstyle=\scriptsize\ttfamily,
  keywordstyle=\bfseries\color{javapurple},
  commentstyle=\itshape\blue,
  identifierstyle=\blue,
  belowskip=1pt,
  frame=shadowbox,
  backgroundcolor=\color{gray01},
  gobble=0
}

\lstdefinestyle{codit}{
  belowcaptionskip=\baselineskip,
  breaklines=true,
  language=java,
  showstringspaces=false,
  basicstyle=\scriptsize\ttfamily,
  keywordstyle=\bfseries\color{javapurple},
  commentstyle=\itshape\blue,
  identifierstyle=\blue,
}

\newtcblisting{mylisting}[2][]{
    arc=3mm,
    listing only, 
    listing style=codit,
    title=#2,
    #1
    }

\newcommand\blue[1]{\textcolor[rgb]{0.00,0.00,1.00}{{#1}}}

\definecolor{blueish}{RGB}{252, 252, 255}
\definecolor{greenish}{RGB}{252, 255, 252}
\definecolor{redish}{RGB}{255, 250, 250}

\definecolor{gray05}{gray}{0.95}
\definecolor{gray08}{gray}{0.92}
\definecolor{gray10}{gray}{0.90}
\definecolor{gray12}{gray}{0.88}
\definecolor{gray15}{gray}{0.85}
\definecolor{gray18}{gray}{0.82}
\definecolor{gray20}{gray}{0.80}
\definecolor{gray25}{gray}{0.75}
\definecolor{gray30}{gray}{0.70}
\definecolor{gray35}{gray}{0.65}
\definecolor{gray40}{gray}{0.60}
\definecolor{gray45}{gray}{0.55}
\definecolor{gray50}{gray}{0.50}
\definecolor{gray55}{gray}{0.45}
\definecolor{gray60}{gray}{0.40}
\definecolor{gray65}{gray}{0.35}
\definecolor{gray70}{gray}{0.30}
\definecolor{gray75}{gray}{0.25}
\definecolor{gray80}{gray}{0.20}
\definecolor{gray85}{gray}{0.15}
\definecolor{gray90}{gray}{0.10}
\definecolor{gray95}{gray}{0.05}

\definecolor{fstarid}{rgb}{0.28,0.07,0.07}

\definecolor{addition}{rgb}{0,0.1,0.5}
\definecolor{dkblue}{rgb}{0,0.0,0.7}
\definecolor{dkgreen}{rgb}{0,0.4,0}
\definecolor{dkred}{rgb}{0.6,0,0}
\definecolor{dkpurple}{rgb}{0.55,0,0.75}
\definecolor{purple}{rgb}{0.69,0,.87}
\definecolor{olive}{rgb}{0.4, 0.4, 0.0}
\definecolor{teal}{rgb}{0.0,0.4,0.4}
\definecolor{azure}{rgb}{0.0, 0.5, 1.0}
\definecolor{gray}{rgb}{0.5, 0.5, 0.5}
\definecolor{dkgrey}{rgb}{0.2, 0.2, 0.2}
\definecolor{lilac}{rgb}{0.70, 0.04, 0.08}
\definecolor{applegreen}{rgb}{0.55, 0.71, 0.0}

\definecolor{step1}{HTML}{CC0066}
\definecolor{step2}{HTML}{CC6600}
\definecolor{step3}{HTML}{440077}
\definecolor{step4}{HTML}{007700}
\definecolor{step5}{HTML}{0000FF}

\xspace%

\newcommand\fstardataset{\textsc{FStarDataSet}\xspace}

\newcommand\fstar{F$^\star$\xspace}

\newcommand\maybecolor[1]{\color{#1}}
\newcommand{\Comment}[1]{}

\newcommand{\nik}[1]{{\todo{Nik:  {\color{violet} #1}}}}

\newcommand{\checkme}[1]{{\color{red}#1}}

\newcommand{\cross}{\ding{55}}
\newcommand{\tick}{\ding{51}}

\newcommand{\citep}[1]{\cite{#1}}

\newtcbox{\inlinebox}[1][]{enhanced,
 box align=base,
 nobeforeafter,
 colback=blueish,
 size=small,
 left=0pt,
 right=0pt,
 boxsep=2pt,
 #1}

\newcommand{\RQ}[2]{%
    {
    \footnotesize
    \vspace{.3em}
    \begin{question}
        \label{rq-#1}
        \noindent\textbf{{RQ{#1}.~#2}}
    \end{question}}
}

\newcommand{\RS}[2]{%
    {
    \begin{result}
        \textbf{\hyperref[rq-#1]{Result {#1}}:~}{{#2}}%
    \end{result}}
}

\renewcommand{\cref}[1]{\Cref{#1}}

\newcommand{\rqa}{Can language models effectively synthesize \fstar definitions given the specification?}

\newcommand{\rqb}{What are the models' success rate across different problem types and what kinds of errors do they produce?}

\newcommand{\rqc}{How do the different prompt components impact the effectiveness of language models?}

\newcommand{\para}[1]{
    \smallskip
    
    \underline{\it #1.}
}

\newcommand{\experiment}{
    \para{Experimental Setup}
}

\newcommand{\resultsec}{
    \para{Results}
}

\newcommand\phitwo{\ensuremath{\mbox{\textit{Phi2}}}}
\newcommand\full{\ensuremath{\phitwo_{\mbox{\textit{full}}}}\xspace}
\newcommand{\fideal}{\ensuremath{\phitwo_{\mbox{\textit{ideal}}}}\xspace}
\newcommand{\nopremise}{\ensuremath{\phitwo_{\mbox{\textit{-pre}}}}\xspace}
\newcommand{\norelated}{\ensuremath{\phitwo_{\mbox{\textit{-re}}}}\xspace}
\newcommand{\nocontext}{\ensuremath{\phitwo_{\mbox{\textit{-ctx}}}}\xspace}

\let\ls\lstinline

\usepackage[turnoff]{notes}

\def\BibTeX{{\rm B\kern-.05em{\sc i\kern-.025em b}\kern-.08em
    T\kern-.1667em\lower.7ex\hbox{E}\kern-.125emX}}
\begin{document}

\title{Towards Neural Synthesis for\\ SMT-Assisted Proof-Oriented Programming}

\author{
    \IEEEauthorblockN{
        Saikat Chakraborty$^\S$, Gabriel Ebner$^\S$, Siddharth Bhat$^\dagger$\IEEEauthorrefmark{1}, Sarah Fakhoury$^\S$,\\Sakina Fatima$^\ddagger$\IEEEauthorrefmark{1}, Shuvendu Lahiri$^\S$, Nikhil Swamy$^\S$
    }
    \IEEEauthorblockA{
        $^\S$Microsoft Research, Redmond, WA, USA \\
        $^\dagger$University of Cambridge, Cambridge, UK, $^\ddagger$University of Ottawa, Ottawa, ON, Canada\\
        {\small \{saikatc, gabrielebner, sfakhoury, shuvendu,  nswamy\}@microsoft.com}\\
    {\small $^\dagger$sb2743@cam.ac.uk, $^\ddagger$sfati077@uottawa.ca}\\
    }
}

\maketitle
\begingroup\renewcommand\thefootnote{$^*$}
\footnotetext{Work done as interns at Microsoft Research}
\endgroup
\begin{abstract}
Proof-oriented programs mix computational content with proofs of program correctness.
However, the human effort involved in programming and proving is still substantial, despite the  use of Satisfiability Modulo Theories (SMT) solvers to automate proofs in languages such as \fstar. 

Seeking to spur research on using AI to automate the construction of proof-oriented programs, we curate a dataset of 600K lines of open-source \fstar programs and proofs, including software used in production systems ranging from Windows and Linux, to Python and Firefox. Our dataset includes around 32K top-level \fstar definitions, each representing a type-directed \emph{program and proof synthesis} problem---producing a definition given a formal specification expressed as an \fstar type. We provide a program-fragment checker that queries \fstar to check the correctness of candidate solutions. We also report on an extended version of our dataset containing a total of 940K lines of programs and proofs, with a total of 54k top-level \fstar definitions. We believe this is the largest corpus of SMT-assisted program proofs coupled with a reproducible program-fragment checker.

Grounded in this dataset, we investigate the use of AI to synthesize programs and their proofs in \fstar, with promising results. Our main finding in that the performance of fine-tuned smaller language models (such as Phi-2 or StarCoder) compare favorably with large language models (such as GPT-4), at a much lower computational cost. We also identify various type-based retrieval augmentation techniques and find that they boost performance significantly. With detailed error analysis and case studies, we identify potential strengths and weaknesses of models and techniques and suggest directions for future improvements.

\end{abstract}

\begin{IEEEkeywords}
Proof Oriented Programming, AI for Proofs, Trustworthy AI programming
\end{IEEEkeywords}

\section{Introduction}
\label{sec:intro}

The recent excitement around AI-assisted programming has been tempered by concerns around the trustworthiness of AI-generated code~\cite{pearce22asleep}. Languages that offer static guarantees can help reduce some of these concerns, e.g., having AI generate safe Rust code rather than C eliminates the risk of AI-introduced memory safety issues. Taking this line of thinking to its limit, using AI to generate code in \emph{proof-oriented languages} which allow programs to be augmented with specification and proofs of correctness could eliminate
trust in AI-generated code, so long as the specification can be audited to match a human's intent. Conversely, proof-oriented languages often require a high-level of human expertise---AI assistance could help make them easier to use.

Recognizing the potential dual benefit (i.e., trustworthy AI programming \& easier program proof) many researchers have begun investigating using AI to synthesize proofs. However, most of the prior work has focused on using AI with tactic-based proof assistants, 
such as Coq, Lean, and Isabelle~\cite{first2023baldur, lample2022hypertree, thakur2023language, sanchez2020generating}, including projects like CoqGym~\cite{yang2019learning}, which builds models based on hundreds of Coq projects containing more than 70,000 tactic scripts, and LeanDojo~\cite{yang2024leandojo}, which uses mathlib~\cite{mathlib20}, a large corpus of formalized mathematics in Lean. AI automation for proof-oriented programming languages like \fstar~\cite{fstar2016}, Dafny~\cite{leino2010dafny},  Viper~\cite{viper16}, Verus~\cite{verus23} and others has received comparatively less attention. The prior work~\cite{kamath2023finding, chakraborty2023ranking, pei2023can,liu2023towards,sun2023clover, rakib2024towards} has been limited by the availability of data, focusing instead on small, hand-crafted problem sets numbering in the few hundreds. This is unfortunate, since proof-oriented languages may be close to the limit of symbolic automation provided by SMT solvers and AI automation could open the door to further automation that has remained out of reach. Additionally, to achieve the promise of trustworthy AI programming, we believe it is essential to develop AI for \emph{program and proof synthesis} rather that only on mostly mathematical tactic-based proofs.

Towards this end, our paper makes the following three major contributions:

\smallskip\noindent{\bfseries\textit{1. A new dataset of programs and proofs:}} Aiming to spur research in AI-assisted proof-oriented programming, our first main contribution is \fstardataset, a dataset of \fstar programs and proofs extracted from {2060} source files, representing about 600K lines of source code, drawn from {8} open source projects. The dataset provides around 32K top-level \fstar definitions, coupled with tooling that allows each definition to be checked in isolation. We believe this is the largest dataset of SMT-assisted proof-oriented programs and we envision for it to be a live, evolving data set, with new projects added to it over time. Indeed, currently \fstardataset has grown to include {4} additional projects reaching 940K lines of source code (see \Cref{v2_data}, and \ref{v2_result} for details and initial results). Although we currently focus on \fstar, we hope for the dataset to also grow to include data from other proof-oriented languages. We describe the dataset in detail in \S\ref{sec:benchmark}.

\smallskip\noindent{\bfseries\textit{2. A benchmark problem:}} Grounded in this data, we design a straightforward benchmark: \emph{given a type as a formal specification, synthesize a program that has that type}. Each of the 32K definitions in \fstardataset yields its own synthesis problem, where the type of the definition is the ``goal" type, and a technique is expected to synthesize a definition that the \fstar compiler attests is goal-type-correct. In \fstar, types are rich enough to capture a variety of specifications, ranging from simple types as in other functional programming languages, to dependently typed specifications that capture functional correctness properties of a program, i.e., types can represent arbitrary propositions. Dually, programs in \fstar contain computational content (e.g., logic for reversing a list), but they can also be interpreted as proofs. As such, our benchmark can be seen as an instance of a type- or specification-directed program \& proof synthesis problem. We present a simple and objective taxonomy to classify benchmark instances, distinguishing simply typed, dependently typed, and fully specified proofs, corresponding roughly to the precision of specifications.

\smallskip\noindent{\bfseries\textit{3. Designing and evaluating neural synthesis techniques:}} We apply a (by now) standard regime of prompting large language models (LLMs) to generate solutions, backed by retrieval augmentation and fine-tuning techniques specific to our setting. In particular, we construct a prompt by augmenting the goal type with related types and definitions from the program context, based on various embedding models. In \S\ref{result}, we evaluate the performance of off-the-shelf large language models, including GPT-3.5~\cite{openai2023gpt} and GPT-4~\cite{openai2023gpt4}, as well as fine-tuned smaller models include Phi2-2.7B~\cite{gunasekar2023textbooks}, Orca2-7B~\cite{mitra2023orca}, and StarCoder-15B~\cite{li2023starcoder}, with the following main takeaways.

\begin{itemize}
\item Fine tuned smaller models can match or outperform large language models (\S\ref{sec:prompt_v_finetune}). 

\item Different classes of problems are solved with varying degrees of success, with common error modes differing between pretrained and fine-tuned models (\S\ref{sec:case_study}).

\item Leveraging the contexts as well as retrieval augmentation significantly boosts the quality of results (\S\ref{sec:input_modalities}).
\end{itemize}

Based on our results, we are optimistic about for the future of AI-assisted proof-oriented programming. Researchers building verified systems in proof-oriented languages have reported writing around 3-5 lines of proof for each line of verified code~\cite{ironclad14osdi,haclstar}, a considerable overhead, despite strong symbolic automation from SMT solvers. For SMT-assisted proof-oriented programs, we provide the first, substantial empirical evidence that LLMs trained on corpora like \fstardataset, and prompted with retrieval augmentation, can automate somewhere between a third and a half of proofs. That said, our approach focuses on synthesizing program and proof fragments given their specifications: finding the right modular decomposition to prove a program correct, with the right specifications and auxiliary lemmas, is not yet within the scope of the techniques we explore. \S\ref{sec:discussion} provides further discussion and analysis.
We release \fstardataset in 
{\tt \url{https://huggingface.co/datasets/microsoft/FStarDataSet}}. 
Details of the project, source code and models can be found in {\tt \url{http://fstar-lang.org/popai}}

\section{Background}
\label{sec:background}

We start by providing some general background on \fstar, adapted from its online manual.
\fstar is a dependently typed programming language and proof assistant. It encourages \emph{proof-oriented programming}, a paradigm in which one co-designs programs and proofs which attest various aspects of a program’s correctness, including its input/output behavior together with its side effects, if any.

The core design philosophy of \fstar is that the type of a program is a specification of its runtime behavior. Many terms can have the same type and the same term can have many types, e.g., \ls`e : int` states that the term or program fragment \ls`e` reduces to an integer; 
and \ls`e : nat` states that \ls`e` reduces to a non-negative integer,
where \ls`nat = x:int{x >= 0}` is a \emph{refinement} of type \ls`int`. When proving a program \ls`e` correct, one specifies the properties one is interested in as a type \ls`t` and then tries to convince \fstar that \ls`e` has type \ls`t`, i.e., deriving \ls`e : t`. 

Many dependently typed languages have the same core design, however \fstar is distinctive in that in addition to several built-in type-checking algorithms, it uses the Z3 SMT solver~\cite{de2008z3} to try to automatically prove various obligations that arise during type-checking. For example, under the hypothesis that \ls`x : even`, proving that \ls`x + 2 : even`, where
\ls`even = x:int{x 

 The concrete syntax of \fstar is based on other languages in the ML family, including OCaml and F\#. Shown below is a recursive implementation of Quick Sort, together with its specification and proof of correctness. The type of \ls`sort` states that for any total order \ls`f` on elements of type \ls`'a`, given an input list \ls`l:list 'a`, \ls`sort` is a total function (i.e., it always terminates) returning a list \ls`m:list 'a` that is sorted according to \ls`f` and where \ls`m` is a permutation of \ls`l`. The predicates like \ls`total_order`, \ls`sorted`, \ls`is_permutation` etc. are other auxiliary definitions in scope. The implementation of \ls`sort` mixes the computational behavior (i.e., partitioning the list based on a pivot, recursively sorting the partitions, combining and returning them) with proof steps that attest to the correctness of the code with respect to its specification.\footnote{
 Many \fstar examples, including the ones shown here, are similar to program proofs in languages like Dafny or Verus. However, \fstar also allows other styles of higher order and dependently typed definitions that are not expressible in other SMT-assisted languages. We refer the reader to the \fstar manual for a full discussion of similarities and differences.
 }

\begin{lstlisting}
let rec sort (f:total_order_t 'a) (l:list 'a)
: Tot (m:list 'a { sorted f m /\ is_permutation l m })
      (decreases (length l))
= match l with
    | [] -> []
    | pivot :: tl ->
        let hi, lo  = partition (f pivot) tl in
        let res = append (sort f lo) (pivot :: sort f hi) in
    (* <proof> *)
        partition_mem_permutation (f pivot) tl;
        append_count lo hi;  append_count hi lo;
        sorted_concat f (sort f lo) (sort f hi) pivot;
        append_count (sort f lo) (sort f hi);
        permutation_app_lemma pivot tl (sort f lo) (sort f hi);
    (* </proof> *)
        res
\end{lstlisting}

For example, the annotation \ls`decreases (length l)` indicates that the recursion terminates because the length of the list input decreases at each recursive call. Additionally, the source lines delimited by \ls`<proof>` comment tags are calls to \fstar \emph{lemmas}, auxiliary definitions that prove certain properties. For instance, the auxiliary lemma \ls`append_count` is shown below. Its type states that every call to \ls`append_count l m x` guarantees the postcondition described in the \ls`ensures` clause; or, equivalently, the type claims the universal property \ls`forall l m x. count x (append l m) = count x l + count x m`. The proof of this property is by induction on the list \ls`l`, expressed as a recursive function.

\begin{lstlisting}
let rec append_count (l m:list 'a) (x:'a)
: Lemma (ensures (count x (append l m) = count x l + count x m))
= match l with
  | [] -> ()
  | hd :: tl -> append_count tl m x
\end{lstlisting}

Program and proof fragments like \ls`sort`, \ls`append_count` etc. each constitute a \emph{top-level definition} in an \fstar program. Each definition yields a type-directed synthesis problem, i.e., given a goal type such as the first two lines of \ls`append_count` shown above, can an LLM generate a type-correct definition. To help the LLM succeed, we aim to augment the goal type with related information, e.g., the definitions of symbols such as \ls`count`, \ls`append` etc. We evaluate the performance of various retrieval augmentation strategies and LLMs on this task.
\section{A Corpus of Proof-Oriented Programs}
\label{sec:benchmark}


\fstardataset is an archive of source code, build artifacts, and metadata assembled from eight~(8) different \fstar-based open source projects on GitHub, summarized  below, with a focus on  libraries that provide secure low-level software and the tools that enable their proofs.

\newcommand\furl[1]{\iffalse {\footnote{\url{#1}}} \fi}
\begin{itemize}[leftmargin=*]
\item FStar\furl{https://github.com/FStarLang/FStar}: The \fstar compiler itself, including its standard libraries and examples. 

\item Karamel\furl{https://github.com/FStarLang/karamel}: A transpiler from a subset of \fstar called Low* to C, including libraries to work with a model of C types and control structures, e.g., for- and while-loops~\cite{lowstar}.

\item EverParse\furl{https://github.com/project-everest/everparse}: A parser generator for binary formats~\cite{everparse}, used in various large scale systems, e.g., the Windows kernel~\cite{everparse3d}

\item HACL*\furl{https://github.com/hacl-star/hacl-star}: A library of verified cryptographic algorithms~\cite{haclstar}, including ValeCrypt~\cite{valefstar}, a library of verified assembly code, as well as EverCrypt, a cryptographic provider~\cite{evercrypt}, including code deployed in Linux, Firefox, and Python.

\item miTLS-F*\furl{https://github.com/project-everest/mitls-fstar}: A partially verified reference implementation of the TLS protocol~\cite{record}.

\item EverQuic-Crypto\furl{https://github.com/project-everest/everquic-crypto}: A verified implementation of header and packet protection for the QUIC protocol~\cite{everquic}.

\item Merkle-tree\furl{https://github.com/hacl-star/merkle-tree}: A verified, incremental Merkle tree, designed for use in Azure CCF, a confidential computing system~\cite{evercrypt}.

\item Steel\furl{https://github.com/FStarLang/steel}: A concurrent separation logic library, with proofs of data structures and concurrency primitives~\cite{steel}.
\end{itemize}

The dataset will be available publicly as an open source repository referencing the other projects as sub-modules, including a given version of \fstar itself. All the projects are built with the same version of \fstar and the Z3 SMT solver, resulting in a single archive with all the {2,060} source files and a build artifact for each of them, i.e., \fstar's \texttt{.checked} files. Each checked file is accompanied by record of metadata about its contents: for each top-level element (\eg, the definition of a function, or a type, or a type signature), the metadata records its dependences, the compiler settings used to verify them, etc. 
 
\para{Reproducibility \& evolution} We aim to strike a balance between reproducibility of the dataset, while also encouraging the data set to grow and change along with projects it references. Referencing the precise commit hashes of all the referenced projects as sub-modules allows any version of the dataset to be fully reproducible. The results reported in this paper focus on \checkme{version 1} of the dataset, a snapshot from \checkme{November 2023}, provided as an anonymous supplement.

\para{A type-checking harness} To enable using the data to validate program-proof synthesis attempts, we provide scripts that enable launching \fstar and initializing it so that it is capable of checking each top-level definition independently. Once initialized in this way, the \fstar process can be repeatedly queried using an interactive JSON-based API to type-check program fragments---in \fstar, type-checking amounts to program verification. 
However, replaying the \fstar type-checker on top-level definitions in the dataset is not perfect, e.g., due to sensitivity to small changes in the search heuristics used by SMT solvers. 
We use the type-checking harness to identify those top-level definitions whose proofs can be checked in isolation, \checkme{finding that \checkme{90\%} of them meet this criterion. In all our experiments, we focus on this re-checkable fragment of the dataset}.

During the development of the harness, we faced similar challenges as reported by LeanDojo~\cite[A.2]{yang2024leandojo}, mainly:
(a) ensuring names are resolved correctly by opening the right namespace in the right order,
(b) ensuring that definitions to be checked cannot refer to their original version in the library, that they cannot use their original definition implicitly via Z3,
(c) verifying that solutions do not use escape hatches such as \ls`admit ()` or \ls`assume` that are intended for interactive use,
and (d) that all files can be loaded in a consistent state even though the bare dataset has clashing file names and conflicting options.
Challenges (b) and (c) are hard to notice during development because they only result in false positives, i.e., the checker incorrectly claiming that a solution is correct.
The biggest development effort was related to efficiently supporting~(b): we added a feature to \fstar that allows the client to partially load a compiled \texttt{.checked} file until we reach the definition to be checked; this ensures that exactly the same definitions and modules are in scope that were accessible to the original definition.
\checkme{After deduplication, removal of auto-generated code, data type definitions, and all proofs that are fully automated by SMT, the dataset contains {32,054} top-level definitions}.


\begin{table}[t]
    \centering
    \caption{\bf Summary statistics of the \fstardataset.}
    \label{tab:data_stat}
    \resizebox{\linewidth}{!}{
    \begin{tabular}{lrrrr}
    \hlineB{2}
     \multirow{2}{*}{\bf Metric} & \multirow{2}{*}{\bf Train} & \multirow{2}{*}{\bf Valid} & \multicolumn{2}{c}{\bf Test}\bigstrut[t]\\
     & & & {\bf Intra-project} & {\bf Cross-project} \bigstrut[b]\\
    \hlineB{2}
    Number of Definitions & 22779 & 1541 & 5965 & 1769 \bigstrut[t]\\
    Number of Projects & 6 & 6 & 6 & 2 \\
    Number of Files & 1216 & 72 & 306 & 126 \bigstrut[b]\\
    \hline
    Avg. num of lines & 8.66 & 13.63 & 11.40 & 7.45 \bigstrut[t]\\
    Avg. num of tokens & 92.16 & 157.26 & 124.32 & 60.32 \bigstrut[b]\\
    \hline
    \# Simply Typed  & 6736 & 434 & 1248 & 149 \bigstrut[t]\\
    \# Dependently Typed & 12047 & 764 & 3111 & 1431 \\
    \# Proofs & 3996 & 343 & 1606 & 189 \bigstrut[b]\\
    \hlineB{2}
    \end{tabular}
    }
\end{table}

\para{Partitions of the dataset} We split \fstardataset into four sets:
\emph{training}, \emph{validation}, \emph{intra-project test}, and 
{\em cross-project test}.
Each source file is in exactly one of these sets. 
The first three sets are taken exclusively from the \fstar,
Karamel, EverParse, HACL*, Merkle-tree, and Steel projects, while the ``{\em cross-project test}'' (we refer as ``{\em cross-project}'' hereafter) exclusively contains definitions from EverQuic-Crypto and miTLS-F*.
The training set is closed under the dependence relation, \ie a
file in the training set also has all its dependencies in the same set. As such,
the training set contains files close to the roots of the
dependence graphs for each project. 
In contrast, files in the intra-project test set (refers to as ``{\em intra-project}'' hereafter) may depend on files from other sets. 
Partitioning according to the dependence relation ensures
that our trained models have not been contaminated with any information
that depends on the test set. 
Files in cross-project set are from projects whose
files are not in any of the other sets.
This allows us to do a 
generalization 
evaluation across projects of the trained models, while again
minimizing the potential for dataset contamination.

\para{A Taxonomy} We classify the definitions in the dataset by
their type, providing three classes corresponding loosely to the theoretical complexity of synthesizing a type-correct definition. 

\begin{itemize}[leftmargin=*]
\item \emph{Simply typed} definitions have simple types such as \ls`int -> int`. Types that include refinements but no dependences are also considered simply typed, e.g., \ls`nat -> nat`. Definitions that are type polymorphic, e.g., \ls`f:('a -> 'b) -> list 'a -> list 'b`, are also considered simply typed. Types in this class could be expressed in many other general-purpose programming languages. Synthesizing definitions in this class is non-trivial---many symbolic synthesis techniques focus exclusively on simple types, while also excluding type polymorphism or non-dependent refinements~\cite{osera15synthesis}. Nevertheless, types in this class may admit many simple solutions. For example, given a goal type of \ls`nat -> nat`, a solution such as \ls`fun _ -> 42` would succeed.

\item \emph{Proof} definitions have types that represent propositions, e.g., the type of \ls`append_count` shown in \S\ref{sec:background}. Such types are so precise that \fstar's logic allows proving that all well-typed terms of a given proposition type are semantically equal. As such, this is the most challenging class, since there is at most one semantically unique solution and as synthesizing a solution amounts to proving a theorem in one go.

\item \emph{Dependently typed} definitions describe all cases whose types are not in the other two classes. Dependent types can encode complex specifications, e.g., the type of \ls`sort` shown in \S\ref{sec:background} is a detailed functional correctness property. Therefore definitions in this class in general have a higher theoretical complexity than the simply typed class. However there can still be multiple semantically different solutions to a dependently typed goal. For example, an implementation of merge sort can be given the same type as \ls`sort`.
\end{itemize}

\Cref{tab:data_stat} shows  some statistics of  \fstardataset version 1, the basis of all experiments and analyses in this paper.




\section{Neural Program \& Proof Synthesis for \fstar}
\label{sec:exp}


\begin{figure}[t]
    \centering
    \includegraphics[width=\linewidth]{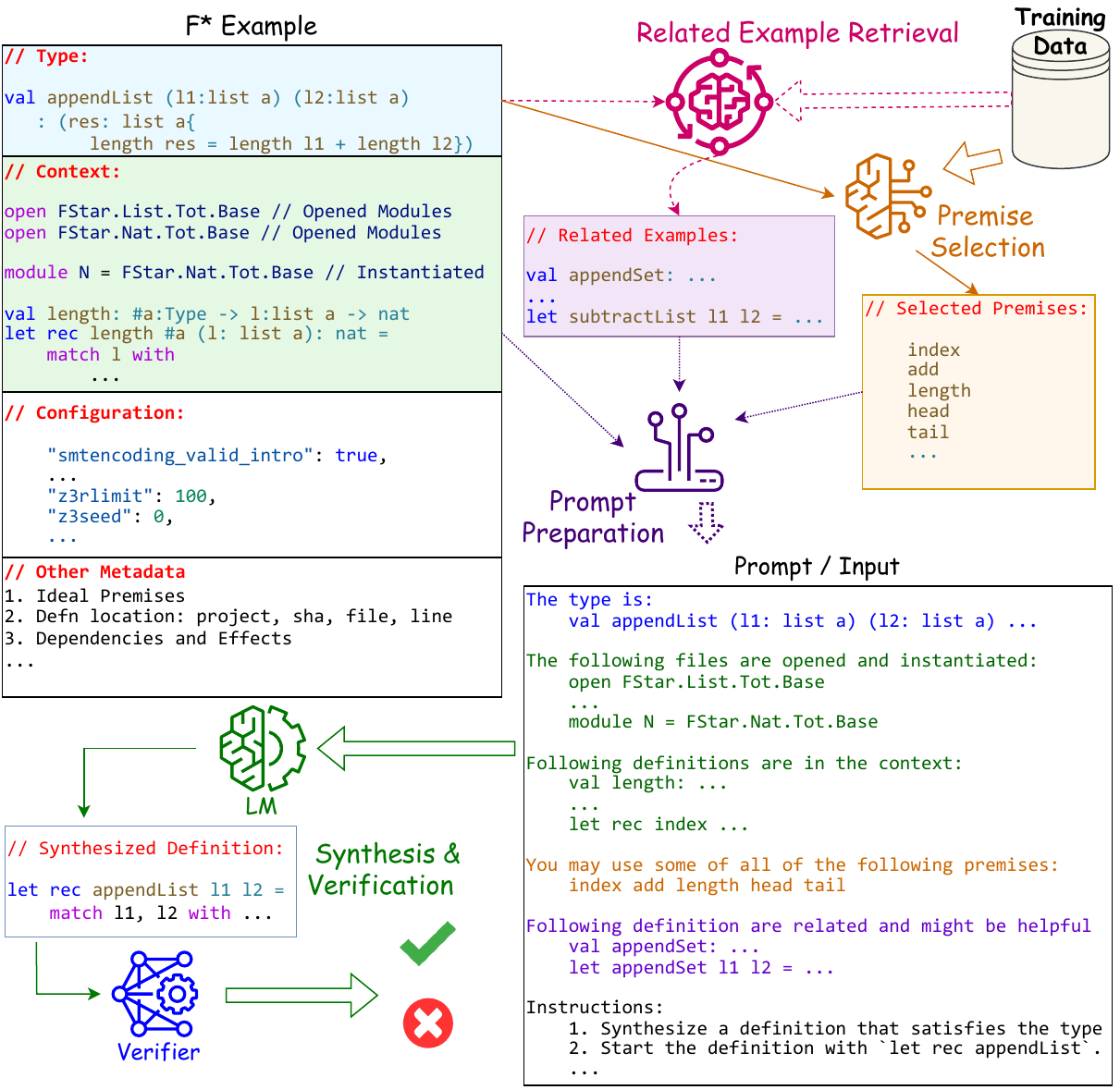}
    \caption{\bf Overview of our experimental approach for synthesizing \fstar definitions.}
    \label{fig:overview}
\end{figure}


\Cref{fig:overview} shows a high level overview of the program and proof synthesis technique we develop in this paper.
The top-left of the figure shows an instance of a top-level definition from \fstardataset, containing the dependent type $t_i$ of a function \ls`appendList`, other metadata, and, importantly, information about its ``file context", i.e., all the top-level elements preceding \ls`appendList` in the file in which it appears. 
To synthesize definitions for $t_i$, first (at top right) we retrieve related examples from the training data and select premises (\ie ingredient definitions) that are likely to be used in the definition body.
At the center of the figure, combining $t_i$ with its file context, related examples, selected premises, we create a prompt $p_i$. We use the prompt and the ground truth definition from the training data ($\mathcal{D}_T$) to train models. Finally, at bottom left, for testing and evaluation, we present the same kind of prompt to our finetuned LLMs, and evaluate the correctness of definitions with the type-checker harness. Note, we do not feedback type-checking results to the LLM for repair, an enhancement we are considering for the future.

\para{Related Examples Retrieval} 
Inspired by successes of providing language models with related examples (colloquially known as Retrieval Augmented Generation, RAG)~\cite{jiang2023active, lewis2020retrieval, parvez2021retrieval}, we provide related examples as input to the model for better generalization. To find related examples, we calculate the similarities between the input type and the types of training examples. Our intuition is that if two examples' types are very close, there will be similarities between their definitions. We compute the similarity between types in their embedding space. Concretely, we embed the types into a fixed dimensional vector space using an LLM-based embedding model, {\tt text-embedding-ada-002}~\cite{neelakantan2022embeddings}, and compute the cosine similarity between $t_i$ and $t_x \in \mathcal{D}_T$. We use the $t_x$ and $d_x$ (\ie definition of $t_x$ from $\mathcal{D}_T$) as related examples.  

\para{Premise Selection}
Following Yang~\etal~\cite{yang2024leandojo}, we also provide the models with other definitions (premises) which are likely to be used in the body of the definitions. We fine-tune an embedding model based on \texttt{all-MiniLM-L6-v2}~\cite{reimers2019sbert}. During inference, given a type as input the fine-tuned premise selection model produces ranked list of premises based on their likelihood of being used in the definition. We also explore fine-tuning alternative embedding models in \cref{subsec:eval_premise}.


To fine-tune an embedding model $e(\cdot),$ we take for every definition $g$ in the training set the definitions $p_1, \dots, p_n$ whose name occurs in the definition's value but not in its type and then randomly sample $q_1, \dots, q_n$ definitions whose name occurs in neither the value nor the type but are in scope at the definition's location. As training objective we use the mean-square error
$\frac{1}{2n} \sum_{i=1}^n
  {(e(g)^T e(q_i))}^2 +
  {(e(g)^T e(p_i) - 1)}^2$,
which aims to align the goal $g$ to the ground-truth premises $p_1,\dots,p_1$,
and making the embedding of the goal orthogonal to the negative premises $q_1,\dots,q_n$.
For evaluation, the premises are then ranked by cosine distance to the goal. 
{Note that, ``Premise Selection'' allowed us to fine-tune model as we already know the ground truth premises from the dataset. In contrast, there is no straightforward notion of ground truth for related example, hence we use embeddings from an LLM that supposedly understands its input. }

\para{Prompt Preparation}
We prepare the input to the LLM (colloquially known as a prompt), by combining the ``goal'' type, the file context, selected premises, retrieved related examples, together with some fixed instructions for the model. We reserve a fixed number of tokens in the prompt for the file context, premises, and related examples. The file context contains opened and instantiated modules in the file together with other definitions in the file up to the point of the goal type, preferring definitions that are closer to the goal to fit in the fixed token budget for the file context. Likewise, for related examples and premises, we rank them according to descending similarity values, including as many as fit into their respective token budgets. We organize different components within a natural language prompt, \ie each component of the prompt is preceded by a brief description of what the component is. 
We represent each of the related example with its type and definition. 
For each of the selected premises, we only use the {\em short name} of that premise in the prompt. While the type and definition of a premise contain more important information about it, we experimentally observe that using other information actually hurt the ultimate performance. Our conjecture is that since the number of allowed tokens for the prompt is limited, including additional information with each premise allows fewer premises to be included in the prompt, which leads to important premises being excluded from the prompt altogether. 
Finally, the prompt ends with a set of instructions including the prefix of the definition. \cref{fig:overview} shows an example prompt at bottom right. We experiment on the impact of different components (context, retrieved examples, and selected premises) in \S\ref{sec:input_modalities}.

\para{Training}
We train different language models of varying sizes to synthesize definitions from prompts. In particular, we trained Phi-2, Orca-2, and StarCoder with 2.7B, 7B, and 15.5B parameters, respectively. The input to the models is
the prompt constructed as described above and the target response is set to be the definition from the dataset itself. We use standard practices for training, including low rank optimization~\cite{hu2021lora}, parameter efficient model tuning~\cite{mao2021unipelt}, and quantized model training, enabling us to deploy the models with GPUs with smaller memory size, for both both training and inference. We train Phi-2 model for 10K steps, Orca-2 model for 5K steps, and StarCoder for 2K steps, requiring roughly similar training time. The training objective is to maximize the probability of the definition given the prompt, which is realized by Cross Entropy Loss.  
We retain the model checkpoint which resulted in the least loss in the validation dataset. 


\para{Synthesis \& Verification}
Having the prompt designed, we synthesize the definition for a given type using the fine-tuned (also pretrained) LMs. While generating, we use temperature based sampling. For each example, we generate $k$ definitions ($S_k$) using the LMs. We evaluate each of these generated definitions using the type-checking harness. In the case of a verification (type check) failure, the checker return the error code with error message. We define the evaluation metric, {\em verify@k} as the percentage of examples in the evaluation data for which there is at least one verified definition in the $k$ generated solutions by the LM.

\section{Empirical Evaluations}
\label{result}

\subsection{Performance of Different Models}
\label{sec:prompt_v_finetune}

In this section, we evaluate the ability of language models to synthesize \fstar programs and proofs,  including of state-of-the-art LLMs such as GPT-3.5 and GPT-4, which are available for inference through APIs, as well as fine-tuned small and medium-sized language models deployed on local machines, posing the following research question:

\RQ{1}{\rqa}

\experiment 
We evaluate all models using the same prompt, as described in \S\ref{sec:exp}. We divide the experiments into two solution strategies: (i) prompting and (ii) fine-tuning. Prompting involves giving a pre-trained language model a task and assessing whether or not it is able to solve it. Fine-tuning, on the other hand, involves further refinement of a model parameters based on a task-specific dataset. For prompting LLMs, we use the OpenAI Python API. For prompting smaller language models, namely Phi-2, Orca-2, and StarCoder, we fine-tune these models (from the HuggingFace library~\cite{wolf2019huggingface}) as described in \S\ref{sec:exp} and use the fine-tuned model checkpoints for inference. We set the same token limits for all models: 500 tokens for context, 400 for related examples, 300 for selected premises, and 500 for generated definitions.

\begin{table}[t]
    \caption{\bf Comparison between Prompting different models and finetuning for synthesizing \fstar definitions. Combined Prompted$^*$ shows the result when we combine the synthesized definitions from all prompted experiments (5), Combined Finetuned$^\dagger$ corresponds to the combined results from finetuned experiments (3), and  Comb. Prompted \& Finetuned$^\ddagger$ shows the results across all the experiments (8).}
    \label{tab:prompting_v_finetuning}
    \centering
    \resizebox{\linewidth}{!}%
    {\begin{tabular}{l|l|r|r|r||r|r|r}
    \hlineB{2}
    \multirow{2}{*}{\bf Strategy} & \multirow{2}{*}{\bf Model} & \multicolumn{3}{c||}{\bf verify@k (intra-project)} & \multicolumn{3}{c}{\bf verify@k (cross-project)} \bigstrut \\
    \cline{3-8}
     & & \textbf{k = 1} &  \textbf{k = 5} & \textbf{k = 10} & \textbf{k = 1} &  \textbf{k = 5} & \textbf{k = 10}  \bigstrut\\
    \hlineB{2}
                & GPT-3.5           &   {13.51}  &     {25.20}   &   {29.81}   &    {6.67}   &      {14.41}   &   {18.54} \bigstrut\\
                & GPT-4             &   {19.41}   &    {31.90}   &   {36.38}   &   {13.00}   &      {23.57}   &   {28.49} \bigstrut \\
    {Prompted}   
                & Phi-2 (2.7B)         &    {0.30}   &       {1.26}   &    {2.20}   &    {0.11}   &        {0.62}   &    {1.75} \bigstrut\\
                & Orca-2 (7B)            &   { 2.63}   &     { 6.39}   &   { 8.55}   &   { 0.45}    &      { 2.20}   &   { 3.84} \bigstrut \\
                & StarCoder (15.5B)     &   { 1.89}   &    { 7.14}   &   {12.27}   &   { 0.73}   &     { 3.79}   &   { 6.90} \bigstrut \\
                \hline
    \multicolumn{2}{c|}{Combined Prompted$^*$} & 26.34 & 40.02 & 45.47 & 16.56 & 29.96 & 36.18\bigstrut\\
                
    \hlineB{2}
    \multirow{4}{*}{Finetuned}   
                & Phi-2 (2.7B)             &   {17.28}   &    {27.75}     &   {31.10}       &    {8.59}       &   {16.62}       &   {20.97} \bigstrut \\
                & Orca-2 (7B)           &   {14.90}   &     {26.02}     &   {30.61}       &   { 7.74}       &     {14.92}       &   {18.37} \bigstrut \\
                & StarCoder (15.5B)         & {\bf 27.95} &     {\bf 39.50} &   {\bf 43.98}   &   {\bf 16.73}   &      {\bf 27.64}   &   {\bf 32.90} \bigstrut \\
    \hline
    \multicolumn{2}{c|}{Combined Finetuned$^\dagger$} & 32.86 & 44.17 & 48.52 & 20.97 & 32.28 & 37.20\bigstrut\\
    \hlineB{2}
    \multicolumn{2}{c|}{Comb. Prompted \& Finetuned$^\ddagger$} & 38.76 & 50.49 & 55.34 & 27.13 & 40.14 & 45.56\bigstrut\\
    \hlineB{2}
    \end{tabular}}

    $^*:$ verify@k interprets as verify@5k.
    $^\dagger$: verify@k interprets as verify@3k, and $^\ddagger$: verify@k interprets as verify@8k.
    
\end{table}

\resultsec \Cref{tab:prompting_v_finetuning} shows the results of different strategies across different models. We focus on the verify@k metric defined in \S\ref{sec:exp}, but also report
a combined verify@$n$k metric which considers an example solved if the instance was solved by \emph{any} of $n$ models under the verify@k metric.

Both GPT-3.5 and GPT-4 solve a significant fraction of problems in the verify@10 metric, though GPT-4 is better. Their success is likely due to \fstar code being part of their training data.
Both models also perform better on intra-project examples than on cross-project examples. When constructing prompts, we only retrieve related examples from the training data split, which specifically does not include any examples from the cross-project set---as such, the cross-project examples benefit less from retrieval augmentation.

The performance of prompted smaller models (Phi-2, Orca-2, and StarCoder) is significantly worse.
Among these models, we do observe a positive correlation between the model size and performance, with the 15.5B parameter StarCoder performing better than Orca-2 (7B parameters), which in turn is better than Phi-2 (2.7B parameters). 
Their relative performance may possibly also be attributed to their pre-training data. Unlike Phi-2 and Orca-2, StarCoder's pre-training data contains OCaml code from GitHub; the syntactic similarity between OCaml and \fstar could also contribute to the relatively higher success of prompted StarCoder.

When we fine-tune these models, their performance significantly improves.
For example, fine tuning Phi-2 improves its verify@10 score by more than a factor of 10.
In fact, fine-tuned Phi-2 slightly outperforms GPT-3.5, while fine-tuned StarCoder also outperforms GPT-4. The combined verify@$n$k results suggest that one could use one or more cheaper, fine-tuned models for synthesis at first, falling back to the larger models only when the smaller models fail.


\begin{figure}[t]
    \centering
    \begin{subfigure}[b]{.45\linewidth}
        \centering
        \includegraphics[width=.9\linewidth]{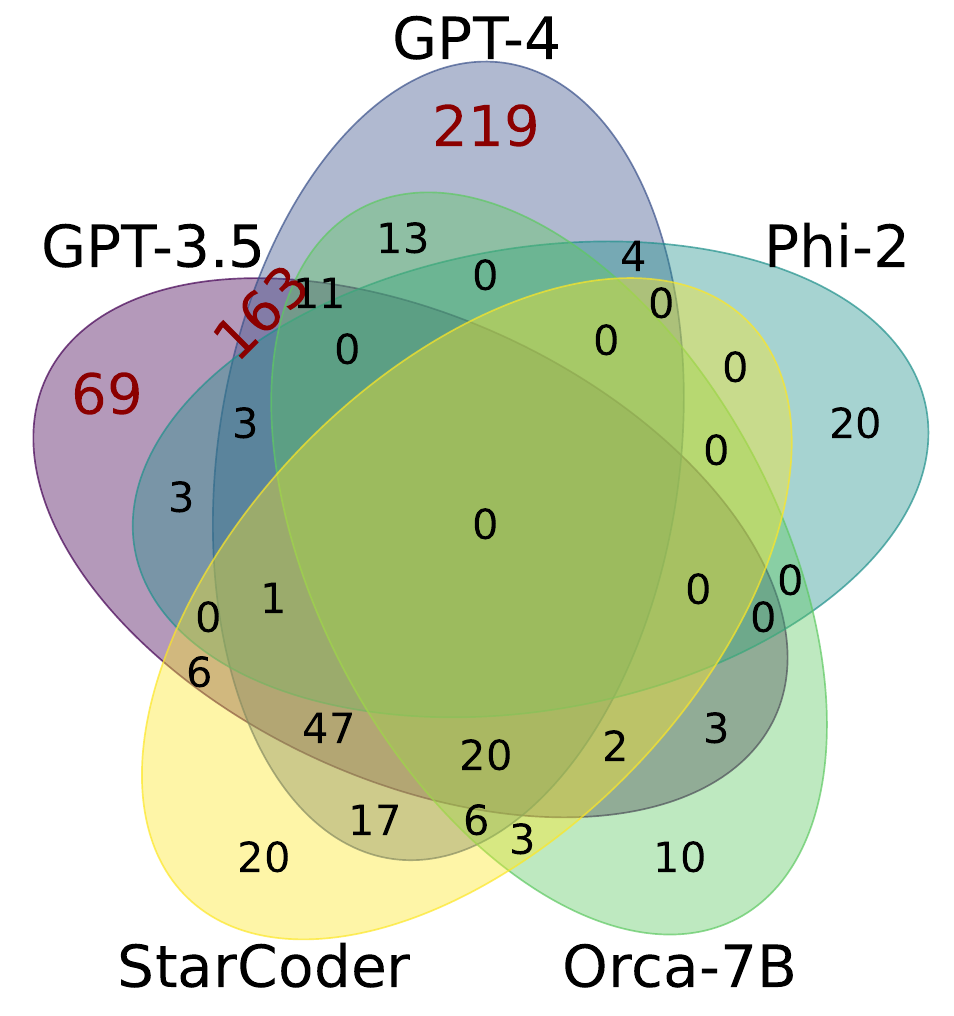}
        \caption{Prompted Models}
        \label{subfig:prompted_venn}
    \end{subfigure}%
    \begin{subfigure}[b]{.45\linewidth}
        \centering
        \includegraphics[width=.9\linewidth]{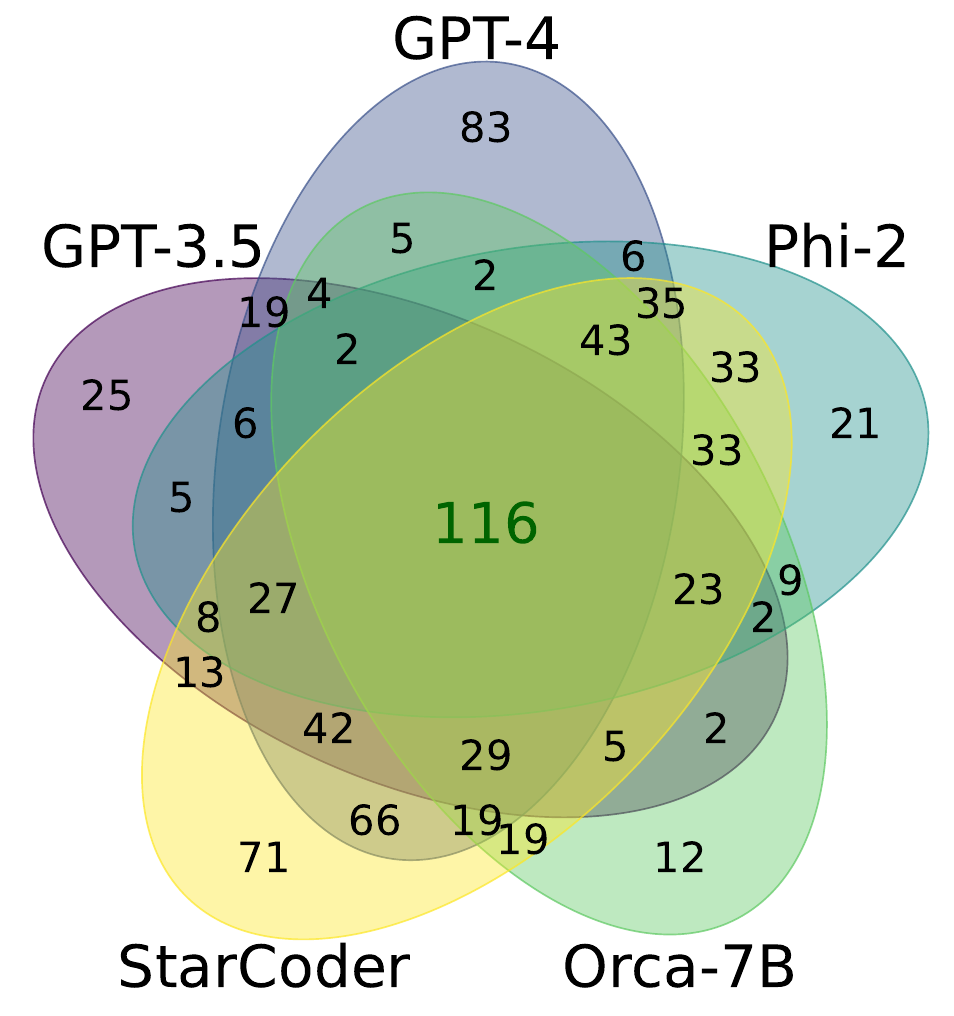}
        \caption{Finetuned Models}
        \label{subfig:finetuned_venn}
    \end{subfigure}
    %
    \caption{\bf Venn Diagram showing intersection of examples from Cross-Project examples solved by different models. The GPT-* performance in Fig. (b) are from prompting respective models. }
    \label{fig:venn_diagram}
\end{figure}

Fine-tuning equips a model with project-specific code idioms to improve performance. 
However, we argue that fine-tuning also yields benefits transferable across projects. 
\cref{fig:venn_diagram} shows the intersection of successes in the \emph{cross-project examples} between different models in both prompted and fine-tuned settings.
As \Cref{subfig:prompted_venn} shows, 451 examples are correctly solved by GPT-3.5 and GPT-4 exclusively, which could not be solved by prompting any other models;  other models exclusively solved only 53 examples. In contrast, after fine-tuning (\cref{subfig:finetuned_venn}), prompted GPT models could only exclusively solve 137 problems, the rest 314 are solved by at least one fine-tuned model. On the other hand, 165 examples are exclusively solved by the smaller models that neither GPT-4 nor GPT-3.5 can solve.
We believe this demonstrates the effectiveness of fine-tuning beyond specific projects they are trained on. 



\RS{1}{Language models are helpful in automating the synthesis of definitions and proofs in \fstar. Fine-tuning smaller models can match or outperform the performance of large language models. Despite being significantly smaller, the fine-tuned Phi-2 (2.7B) model slightly outperforms GPT-3.5 by up to 5\%. In addition, StarCoder (15.5B) outperforms the most advanced GPT-4 by up to 21\%. Further, our evaluation shows the generalizability of these fine-tuned models on unseen projects.}

\subsection{Understanding models' successes and failures}
\label{sec:case_study}

In this section we focus on the following research question:

\RQ{2}{\rqb}

\experiment From \S\ref{sec:benchmark}, recall that we divide the problems in our dataset into three categories, \ie (i) simply typed definitions, (ii) dependently typed definitions, and (iii) proofs. We report on the performance of models in each of these categories. We also analyze
the types of errors reported by the type-checking harness on synthesized code. We divide the errors into three broad categories (i) Syntax errors: where the \fstar could not parse the generated code, (ii) Identifier not found errors: where the definition was parsed, but one or more identifiers could not be resolved, and (iii) Semantic errors: the definition is syntactically valid and has no unresolved identifiers, but the type-checker could not verify the code against the goal type. 

\begin{figure*}[t!]
    \centering

    \begin{subfigure}[t]{0.50\linewidth}
        \centering
        \includegraphics[width=0.95\linewidth]{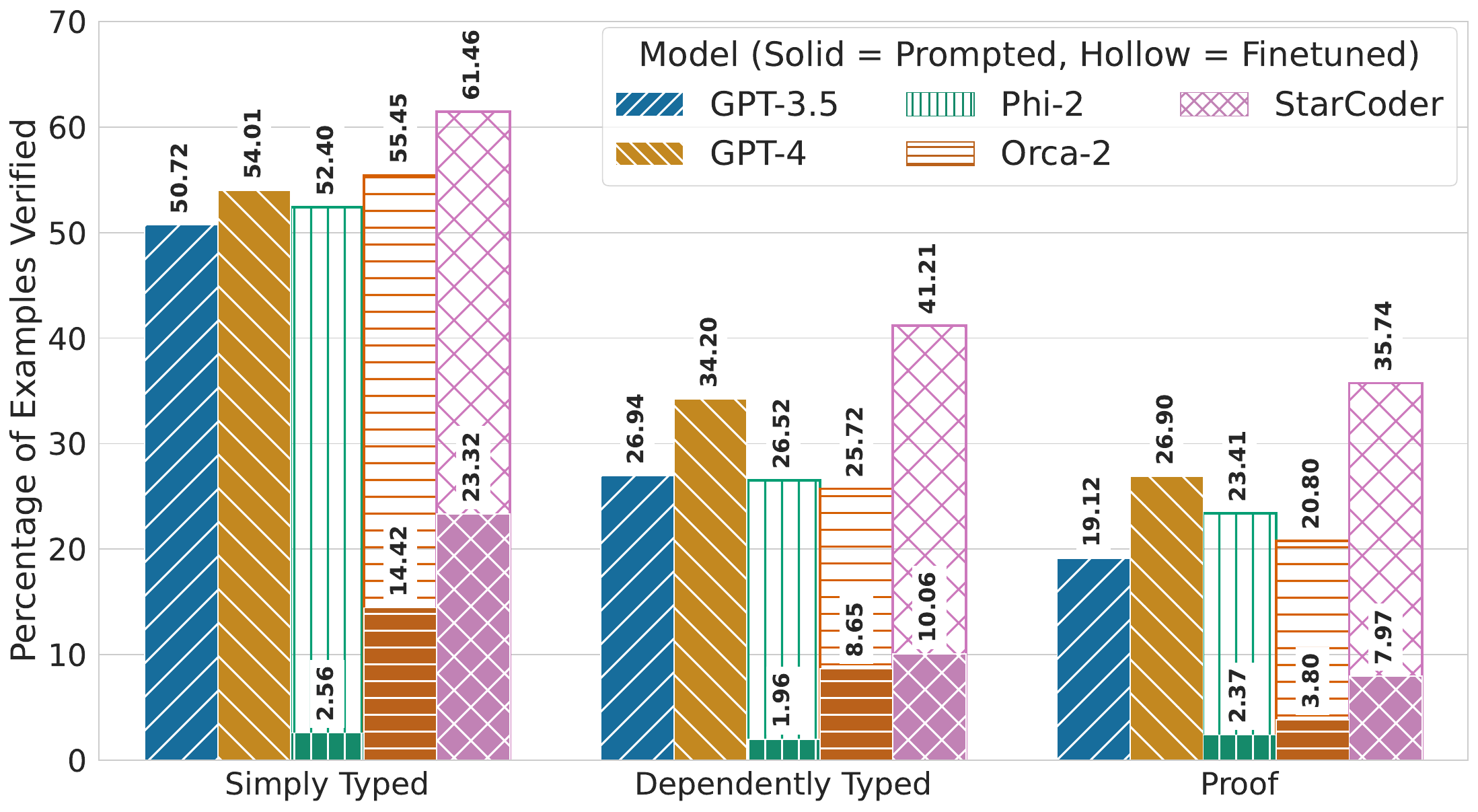}
        \caption{Intra Project evaluation}
        \label{subfig:rq1_class_test}
    \end{subfigure}%
    \begin{subfigure}[t]{0.50\linewidth}
        \centering
        \includegraphics[width=0.95\linewidth]{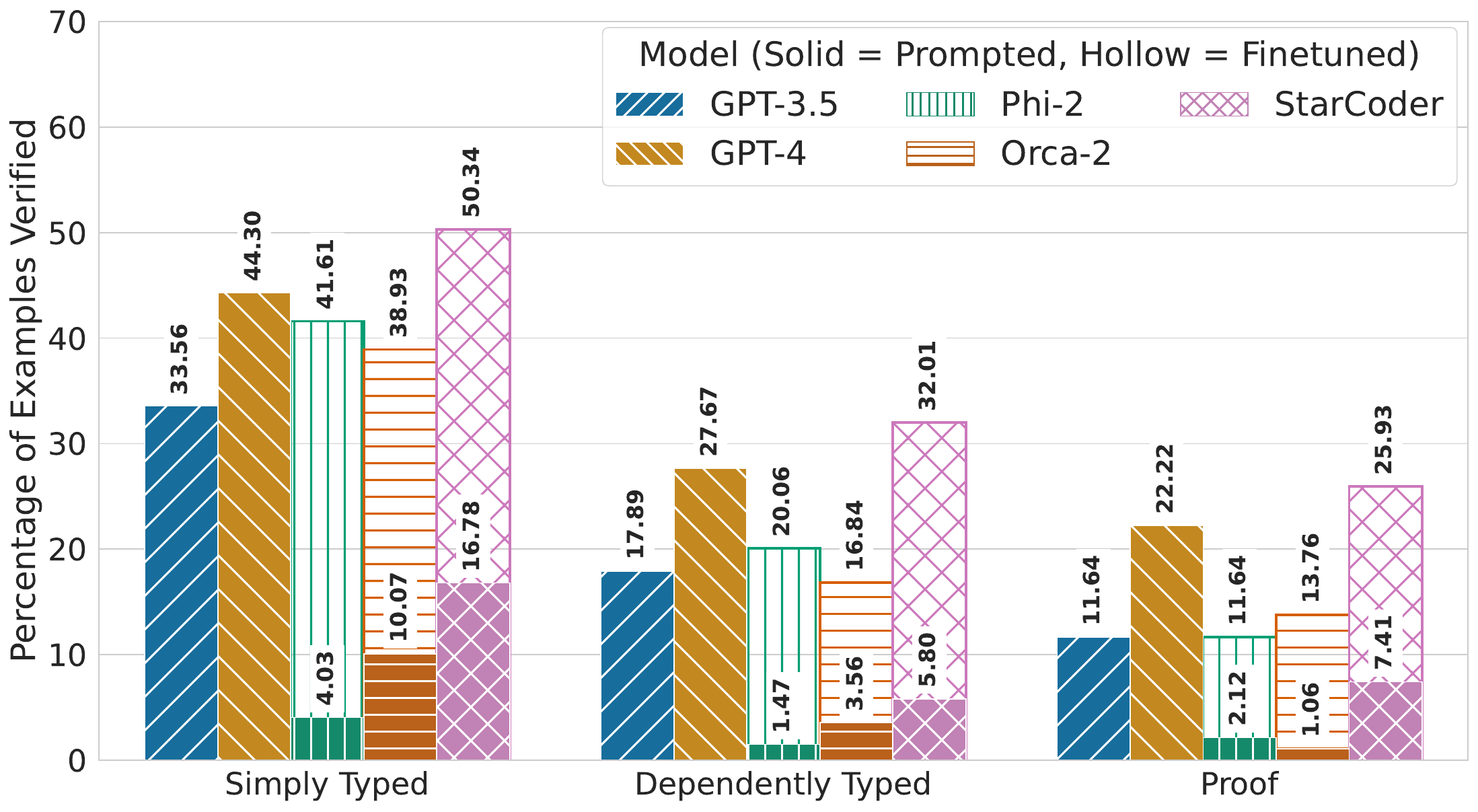}
        \caption{Cross Project evaluation}
        \label{subfig:rq1_class_challenging}
    \end{subfigure}
    \caption{\bf Verify@10 across different types of examples for different models.}
    \label{fig:class_acc}
\end{figure*}

\begin{figure}[h]
    \centering
    \includegraphics[width=.92\linewidth]{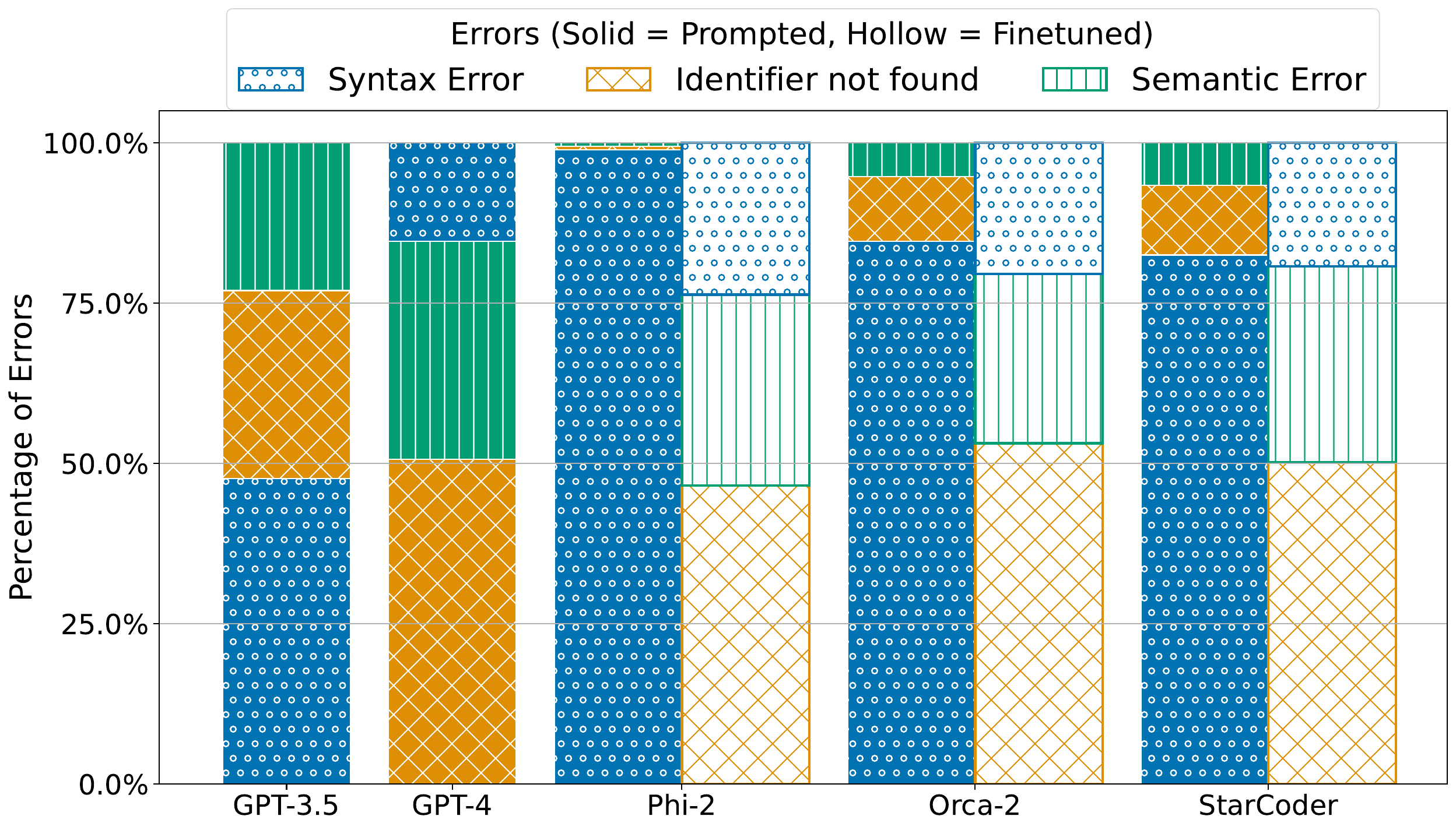}
    \caption{\bf Different errors spawned from the generated definitions by different models, presented as percentage of all errors. }
    \label{fig:model_errors_3_cat}
\end{figure}

\resultsec
\Cref{fig:class_acc} shows the accuracy of different models in different classes of problems under the verify@10 metric. Across all configurations, models are most successful at simply typed definitions. For dependently typed definitions, we observe slightly lower performance across all models compared to simply typed definitions and performance on proof definitions is lower still. This is in keeping with our expectations: as described in \S\ref{sec:benchmark}, our taxonomy is meant to roughly capture the theoretical complexity of the problems and the model performance appears to validate this view. As in \S\ref{sec:prompt_v_finetune}, performance on intra-project examples across all categories is also systematically better than on cross-project examples. In all cases, fine-tuned StarCoder performs the best.


We also examined specific successes from fine-tuned StarCoder to give a qualitative sense of the kinds of problems it is able to solve. We give two representative examples.

Our first example is dependently typed, for \ls`FStar.OrdSet.where`, a function that filters and ordered set \ls`s` to those elements that satisfy a condition \ls`c`. The specification provided in the prompt is very detailed and includes functional correctness. The model is able to synthesize the type-correct definition shown, a fairly typical functional program. To put this in perspective, compared to AI automation for tactic-based proofs, our work exploits an LLMs ability to generate \emph{programs}, and relies on \fstar's dependent types and SMT-based automation to certify synthesis results. 

\begin{lstlisting}[escapechar='']
let rec where #a #f (s:ordset a f) (c:condition a)
  : Pure (ordset a f)
       (requires True)
       (ensures fun (z:ordset a f) ->
           (as_list #a  z == FStar.List.Tot.Base.filter c 
            (as_list s)) /\ (forall x. mem x z = (mem x s && c x)) /\
           (if size z > 0 && size s > 0 then 
            f (head s) (head z) else true))
  = match s with
    | [] -> empty
    | x::q -> 
      let z = where q c in
      if c x then insert' x z else z
\end{lstlisting}

Our next example is from EverQuic-Crypto, a cross-project proof problem. The lemma is about the property of a binary
formatted header, using functions from the EverParse library. The preceding file context for this example contains a similar proof and the related examples includes an example from the EverParse library. In this case, the model has successfully adapted a related proof from the file context. Indeed, program proofs often contain many similar elements, tedious for a human to write, but models are adept at identifying common patterns and adapting them accordingly.
\begin{lstlisting}
let serialize_header_is_retry
  (short_dcid_len: short_dcid_len_t)
  (h: header' short_dcid_len)
: Lemma (
    let s = LP.serialize (serialize_header short_dcid_len) h in
    Seq.length s > 0 /\ 
    (is_retry h <==> 
        (LPB.get_bitfield (U8.v (Seq.index s 0)) 7 8 == 1 /\
        LPB.get_bitfield (U8.v (Seq.index s 0)) 4 6 == 3)
    )
) = serialize_header_eq short_dcid_len h;
  let tg = first_byte_of_header short_dcid_len h in
  let x = LPB.synth_bitsum'_recip first_byte tg in
  LP.serialize_u8_spec x;
  let s = LP.serialize (serialize_header short_dcid_len) h in
  assert (Seq.index s 0 == x);
  assert (is_retry h <==> (
    LPB.get_bitfield (U8.v (Seq.index s 0)) 7 8 == 1 /\
    LPB.get_bitfield (U8.v (Seq.index s 0)) 4 6 == 3
  ))
\end{lstlisting}




Turning our attention to errors, \Cref{fig:model_errors_3_cat} plots three different class of errors made by models as a percentage of the total erroneous solutions they generated. Interestingly, for GPT-3.5, the largest error class is ``Syntax error", whereas for GPT-4 it is ``Identifier not found"---GPT-4 seems to be better at \fstar syntax than GPT-3.5. For the smaller models, when we prompt them, most of the errors are syntax errors. This is not surprising, since none of them have been trained on \fstar. When we fine-tune the models, the largest error class is ``Identifier not found", indicating that models often hallucinate identifiers. For example, the following definition is generated by StarCoder:

\begin{lstlisting}[escapechar='']
let eqList_ok (#a: Type) (d: deq a) : 
    Lemma (decides_eq #(list a) (Raw.eqList d.raw.eq)) =
    let open FStar.Classical in
    let lemma_eq (xs ys: list a) 
        : Lemma (Raw.eqList  d.raw.eq  xs  ys) =
            FStar.List.Tot.lemma_eq_intro (
                Raw.eqList  d.raw.eq
            ) xs ys; () in
    Classical.forall_intro (lemma_eq)
\end{lstlisting}

While seemingly syntactically correct, this definition uses the symbol \ls`lemma_eq_intro`, which is not in scope. Adapting recent techniques to guide models based on lightweight static analyses (e.g., based on the identifiers in scope) is a promising direction for the future~\cite{agrawal2024monitor, wei2023copiloting}.

The last, but not the least, category of error is Semantic error, which is a collection of many different errors. Notable among these include {\tt Type Error}: a value or identifier with incompatible type is used and {\tt Z3 Solver Error}: Z3 cannot prove the SMT query. Program repair techniques, both search-based~\cite{le2011genprog} and LLM-assisted~\cite{first2023baldur}, may help reduce some of these errors, another direction for the future.


\RS{2}{Model performance follows our taxonomy, with simply typed problems being most commonly solved, then dependently typed, and finally proofs. Models are able to generate type-correct functional programs with complex dependent types, and to generate proofs by adapting similar examples. For most of the pre-trained models, the major source of errors are syntax errors, while after fine-tuning, ``Identifier not found" consist of the majority of errors, followed by semantic errors.}

\subsection{Impact of components of the prompt}
\label{sec:input_modalities}

\begin{table}[t]
    \caption{\bf Impact of different sources of information in the finetuned Phi-2 model.}
    \label{tab:finetuning_ablation}
    \centering
    \begin{tabular}{l|c|c|c|c|c}
    \hlineB{2}
    \textbf{Experiment} & \multicolumn{3}{c|}{\textbf{Auxiliary Information}} & \multicolumn{2}{c}{\textbf{verify@10}} \bigstrut[t]\\
    \textbf{Name} & \textbf{Context} & \textbf{RE} & \textbf{Premise} & \textbf{Intra} & \textbf{Cross} \bigstrut[b]\\
    \hlineB{2}
    \full       & \tick & \tick & \tick    &       {31.10}     &       {20.97}       \bigstrut\\
    \hline
    \nocontext  & \cross & \tick & \tick   &       {21.89}     &       {10.97}      \bigstrut[t] \\ 
    \norelated  & \tick & \cross & \tick   &       {25.92}     &       {21.82}       \\
    \nopremise  & \tick & \tick & \cross      &       {31.15}     &       {20.35}       \bigstrut[b]\\ 
    \hline
    \fideal     & \tick & \tick & Ideal       &       {35.84}     &       {23.74}       \bigstrut\\ 
    \hlineB{2}
    \end{tabular}

    \smallskip
    RE = Related example definition from training set
\end{table}

Recall from \S\ref{sec:exp}, a prompt contains the (1) local file context; (2) related examples retrieved from the training data; and, (3) selected premises by the premise selection model. In this section, we evaluate the impact of each of these three components on model performance.

\RQ{3}{\rqc}

\experiment We take as a baseline for our experience the performance a fine-tuned Phi-2 model (\full) that has access to all three prompt components. We fine-tune three other versions of Phi-2, each time dropping one of three three prompt components: for the \nopremise model, we drop the selected premises from the prompt; \nocontext, drops file context information; and \norelated, drops the related examples. In addition, since we know the ideal premises used in the ground truth definition, we fine-tune another version of \full, where instead of the selected premised, we used the actual premises (\fideal). \fideal is not realistic, but it serves as a roof-line for premise selection. We evaluate these variants on verify@10 as well on the numbers of errors each of these models produce. We also experiment with different ways to combine and present these prompt components to the model. Note that, since these experiments entail a large number of fine-tuning runs, we chose to focus on the Phi-2 model, as it is the most cost-effective. We believe our findings should generalize to other models.

\resultsec
\cref{tab:finetuning_ablation} summarizes our results. The baseline \full model performance is as in \cref{tab:prompting_v_finetuning}. When we remove the related examples from the prompt (\norelated) and fine-tune a model, for intra-project, the performance drops by $\sim$5 percentage points. Interestingly, without the related examples, the performance in cross-project examples increases slightly. Since we extract the related examples from other projects in the training set, and there is little to no helpful examples in the \full model for the cross-project examples. Thus we conjecture, the related examples from the same project helps the model most. 

On the other hand, in \nocontext, we observe the performance dropping significantly from the \full model for both intra-project and cross-project examples. Such a drop is not surprising, since even for a human user, the file context has the most relevant information about the target definition, e.g., in the \ls`FStar.OrdSet.where` example shown in \S\ref{sec:case_study}, the local file context \emph{defines} the type \ls`ordset a f` as an ordered list, crucial information for synthesizing a solution.




Finally, when we remove the selected premises (\nopremise), the performance remains very close to the \full model. This is surprising, since prior work~\cite{yang2024leandojo} demonstrates that premise selection improves the performance of proof synthesis. To investigate further, we fine-tuned \fideal, where instead of using premise selection model, we use the ideal premises. The results reveal that, with ideal premises, the performance does improve significantly from \full model to
35.84\% in intra-project and 23.74\%in cross-project. Additionally,
the number of ``Identifier not found" errors with \fideal reduces significantly to 22149 from 30108 such errors in \full. As such,  \fideal suggests that a premise selection model that better approximates the ideal premises can provide a significant improvement---see \S\ref{sec:discussion} for more discussion.

\RS{3}{File context and related examples are very important sources of information for synthesizing definitions in \fstar. Without the context, the performance of a model drops down by up to 29.6\% and up to 16.55\% without the related examples. An ideal premise selection roof line suggests that it may also provide significant improvements, however our current premise selection model does not significantly impact performance.
}

\section{Further Analysis \& Discussion}
\label{sec:discussion}

This section takes a closer look at some of the results from \S\ref{result}. We report on additional experiments and explore potential for future improvements. 

\subsection{Syntactic Evaluation of Model Generated Definitions}

\begin{figure}[t]
    \centering

    \includegraphics[width=0.90\linewidth]{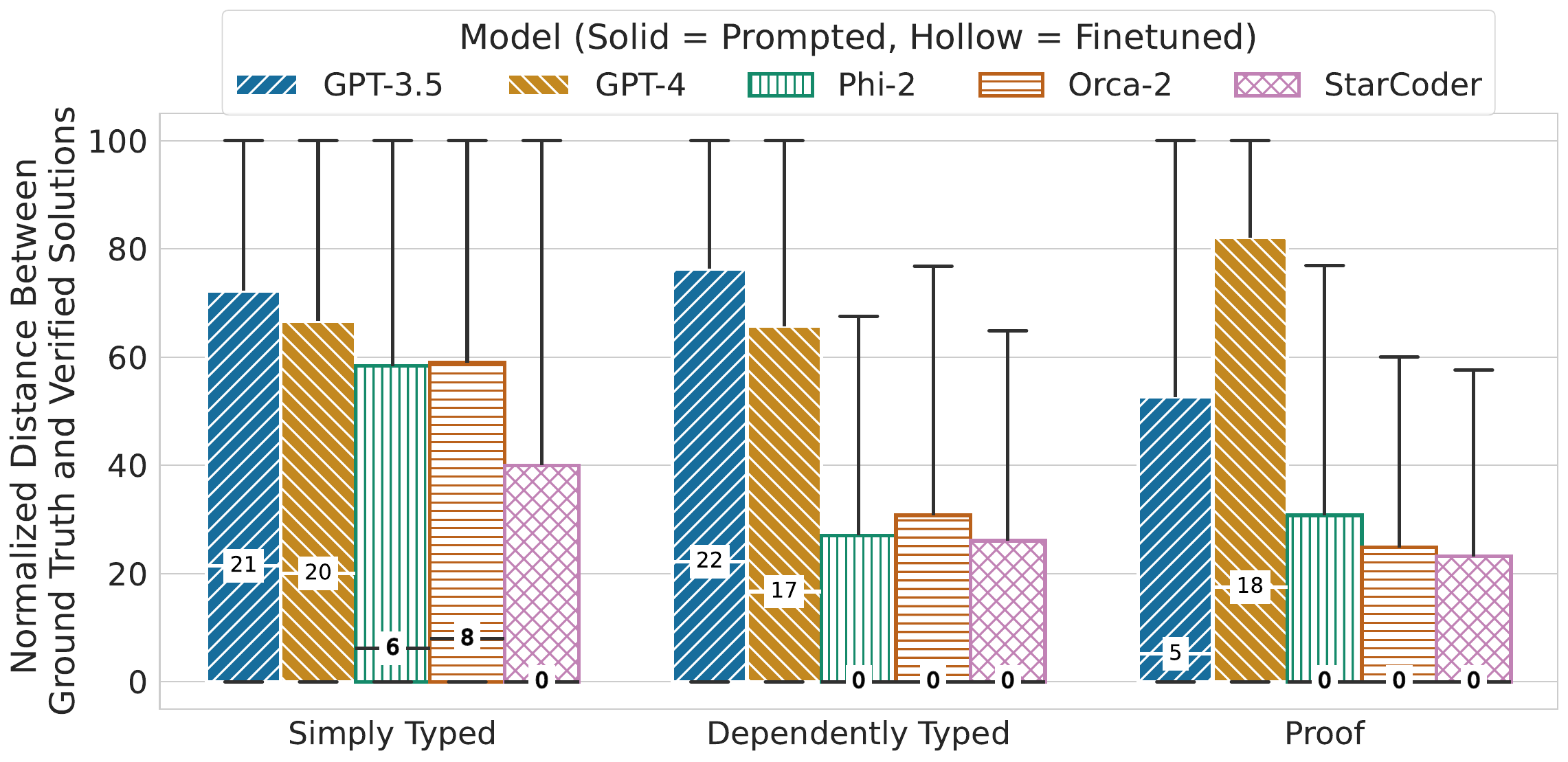}
    \caption{\bf Normalized Levenshtein distance between verified solutions and ground truth across different classes. }
    \label{fig:em_scores}
\end{figure}

\newcommand\levdist{\ensuremath{\mbox{\textit{lev\_dist}}}}
In addition to the semantic verify@k metric of the prior section, we also evaluate how close a model's generated solutions are to the ground truth solution ($g$). We calculate the normalized edit (Levenshtein) distance between tokenized $g$ and solution $v$.  For every problem which has a non-empty set of verified candidate solution $(VS)$, we define the distance as $\min_{v\in VS} \levdist({g}, {v})$. The $\levdist$ calculates the edit distance between $g$ and $v$ and normalized it \wrt the number of tokens in $g$.  \Cref{fig:em_scores} shows the distribution of distances. Prompted LLMs such as GPT-3.5 and GPT-4 generate solutions that are further away from the ground truth compared to fine-tuned models. 

\subsection{Impact of maximum tokens on model performances}

\begin{table}[h]
    \centering
    \caption{\bf Performance of different state-of-the-art LLMs with increasing model capacity for 100 randomly sampled examples.}
    \label{tab:sampled_result}
    \begin{tabular}{l|c|r}
    \hlineB{2}
    {\bf Model} & {\bf Max number of tokens} & {\bf verify@10} \bigstrut\\
    \hlineB{2}
    \multirow{2}{*}{GPT-3.5} & 2,048 & 29 \bigstrut[t]\\
            & 16,000 & 35 \bigstrut[b] \\
    \hline
    \multirow{2}{*}{GPT-4} & 2,048 & 37 \bigstrut[t]\\
            & 16,000 & 50 \bigstrut[b] \\
    \hline
    StarCoder (finetuned) & 2,048 & 45 \bigstrut\\
    \hlineB{2}
    \end{tabular}
\end{table}

Throughout the paper, for a fair comparison, we restrict the maximum number of tokens for different models to 2048 tokens. However, some LLMs accommodate much larger prompts, \eg~{\tt gpt-3.5-turbo-16K} with 16K. While experimenting on these larger context models are expensive, we evaluate GPT-3.5 and GPT-4 models allowing maximum 10K tokens for context, 3K for related examples, and 2K for selected premises, and 1K maximum generation length. To keep the cost of experiment tractable, we randomly sample 100 problems for evaluation. As \Cref{tab:sampled_result} shows, we observe that, with increased capacity, both GPT-3.5 and GPT-4 improves by a large margin compared smaller capacities. While GPT-4 with 16K token capacity achieves 50\% verify@10, fine-tuned StarCoder achieves 45\% with only 2K tokens. Further experiments with GPT-4 on the entire data set would be more definitive, though the cost is prohibitive.

\subsection{Definition ingredients in the prompt and its impact}
\begin{figure}[t]
    \centering
    \includegraphics[width=.90\linewidth]{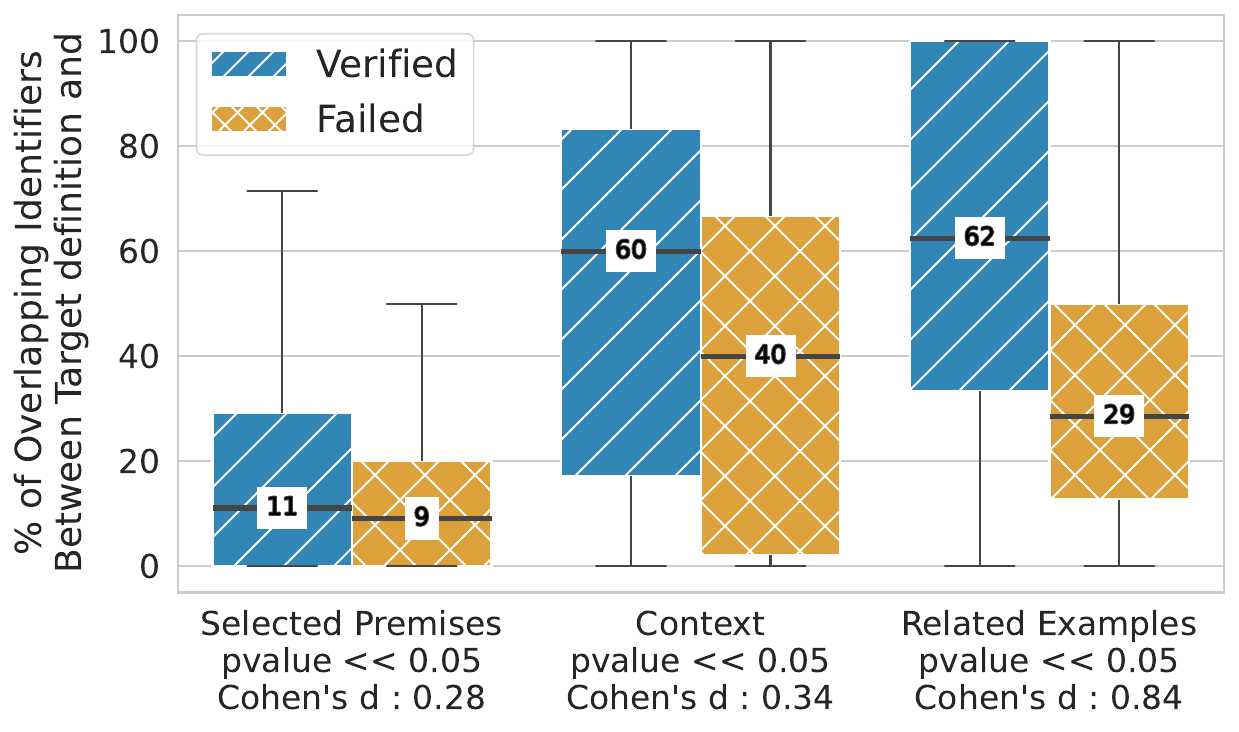}
    \caption{\bf Overlap between identifiers in ground truth and identifiers in each prompt component}
    \label{fig:modality_similarity}
\end{figure}

Following our observation that that majority of the errors are due to hallucinated identifiers, we investigated  how much each of the prompt components help the model generate correct identifiers. We calculated the percentage of overlap between the identifiers used in the ground truth and different prompt components and investigate whether this correlates with model performance. \Cref{fig:modality_similarity} shows that when there is a higher overlap between an information modality and the ground truth identifiers, the models are more likely to generate verified definitions (with statistical significance). While in theory, it is impossible to know the name of the identifiers that will be used in a definition a priori, the analysis suggests that better retrieval augmentation could boost performance. 

\subsection{Evaluating Different Premise Selection Models.}
\label{subsec:eval_premise}

\begin{table}[t]
\centering
\caption{%
\bf Evaluation of alternative models
for the premise selection.
An asterisk (*) indicates a fine-tuned model.
}
\label{tblpremselres}
{
\begin{tabular}{l|cc|cc}
\hlineB{2}
\multirow{2}{*}{\bf Model} & \multicolumn{2}{c|}{\textbf{Intra-Project}} & \multicolumn{2}{c}{\textbf{Cross-Project}} \bigstrut[t]\\
& {\bf MAP} & {\bf NDCG}
& {\bf MAP} & {\bf NDCG} \bigstrut[b]\\
\hline
Pythia 70M & 0.15 & 0.37 & 0.14 & 0.37 \bigstrut[t]\\
Pythia 160M & 0.15 & 0.38 & 0.13 & 0.37 \\
all-MiniLM-L6-v2 & 0.15 & 0.38 & 0.16 & 0.40 \\
OpenAI Ada & 0.19 & 0.42 & 0.19 & 0.43 \bigstrut[b]\\
\hline
Pythia 70M* & 0.33 & 0.55 & 0.15 & 0.40 \bigstrut[t]\\
Pythia 160M* & 0.34 & 0.56 & 0.13 & 0.38 \\
all-MiniLM-L6-v2* & 0.32 & 0.54 & 0.18 & 0.43 \bigstrut[b]\\
\hlineB{2}
\end{tabular}
}
\end{table}

We compared four embedding models for premise selection, before and after fine-tuning: the small 22M parameter model \texttt{all-MiniLM-L6-v2} from {sentence transformers}~\cite{reimers2019sbert}, the \texttt{text-embedding-ada-002} model~\cite{neelakantan2022embeddings} from OpenAI, as well as two generic transformer models from the Pythia~\cite{biderman2023pythia} family with 70M and 160M parameters. 
Comparing the Mean Average Precision (MAP) and Normalized Cumulative Discounted Gain (NDCG) (\Cref{tblpremselres}), without fine-tuning the Ada model achieves roughly 25\% higher performance. Surprisingly, the off-the-shelf Pythia models are competitive with \texttt{all-MiniLM-L6-v2} even though they are not trained on any contrastive objective. After fine-tuning, all of the models have comparable performance irrespective of model size. However, the training does not generalize well and fine-tuning only barely improves performance on the \emph{cross-project} set.

\subsection{Alternative Prompt Preparation}
\label{sec:alternate_prompt}

\begin{table}[h]
    \caption{\bf Different Input Formatting.}
    \label{tab:input_format}
    \centering
    \begin{tabular}{l||c|c}
    \hlineB{2}
    \textbf{Input Type} & \multicolumn{2}{c}{\textbf{verify@10}} \bigstrut[t]\\
     & \textbf{Intra-Project} & \textbf{Cross-Project} \bigstrut[b]\\
    \hlineB{2}
    NL Prompt         &       {\bf 31.10}     &       {\bf 20.97}       \bigstrut[t]\\
    Structured        &       {30.76}     &       {19.45}       \\ 
    Completion        &       {10.54}     &       {12.21}       \bigstrut[b]\\  
    
    \hlineB{2}
    \end{tabular}
\end{table}
In addition to presenting a natural language prompt (NL prompt) (see \S\ref{sec:exp} for details), we investigated alternative prompting strategies. In particular, we created a structured format format similar to the format proposed by Gupta~\etal~\cite{gupta2023grace}. In this version, every input modalities are surrounded by tags such as {\tt <context>} and {\tt </context>}, surrounding the file context  {\tt <related>} and {\tt </related>} around the related examples. In another version, we create a 
completion style where all the prompt components are concatenated with a separator token {\tt <|end\_of\_text|>}, the default padding token of Phi-2, without any description of each of the components. Natural language prompts are slightly better that structured prompts, while the completion-style prompt performs substantially worse. We conjecture that since the Phi-2 model is pre-trained mostly on natural language, wrapping the prompt components with natural language help during fine-tuning; the structured tag formats also help the model distinguish the components. Such a distinction is lost in the completion style prompt. 
\section{Threats \& Limitations}
\label{sec:threats}

\para{Data contamination}
The training data for GPT-3.5 and GPT-4 is not publicly known,
but it is very likely that it intersects the intra-project and cross-project test sets because those are taken from repositories publicly hosted on GitHub.
On the other hand, despite maintaining file level separation between train and test sets, we observe that there are a small number of clones (\ie examples with different names, yet same definitions) between train and test set. In particular, we identified 343 such clones in intra-project test set and 7 in the cross-project test set. 
While these solutions are clones, the context they are in are different. Hence, the prompt to the language model (see \cref{sec:exp} for prompt preparation) are different for these. While the related examples retrieval method are in general capable of finding these clones, we argue that such a setup is not unrealistic for real development, where developer often reuse code written elsewhere. 


\para{Hints about the definition problem statement}
The synthesis problem in this paper requires the model to synthesize a definition for a given type. 
However, there are some implicit hints about the definition that are present in the types, and context. For instance, consider the example \ls`sort` in \S\ref{sec:background}, the type contains an effect \ls`decreases (length l)`, which implicitly tells the model that there will be an induction on the length of the\ls`l`. While these are hints potentially helping the model, we argue that these are the hints that a developer writing \fstar code may already know. 


\para{Preciseness of specifications}
We consider a problem to be solved if the solution type-checks. For problems in the proof class, all solutions are equivalent, so no further inspection is necessary for a successful proof. However, in other cases, specifications may only be partial and, although type-correct, a solution may still be subject to inspection to confirm if it matches a user's under-specified intent. 

\para{Non-deterministic nature of LLMs}
Large Language Models, especially those for which we do not have access to the model weights (\eg GPT-3.5/GPT-4) often go through regular updates. Hence, it it very difficult, if not impossible, to reproduce some of the results from those LLMs. In contrast, the fine-tuned model, being smaller in size, will be accessible to those seeking to reproduce our results. 


\section{Related Work}
\label{sec:related}

In the past 15 years, an increasing number of software systems have been proven correct, both using interactive theorem provers like Coq and Isabelle/HOL, as well using SMT-assisted proof-oriented languages like \fstar and Dafny. Software proofs require expertise and can be quite verbose. A common metric used to evaluate proof effort is a proof:code ratio, i.e., the number of lines of proof for each line of executable code. This ratio is typically higher in interactive proof assistants than in SMT-assisted proof-oriented languages.
For example, sel4 in Isabelle/HOL~\cite{sel4sosp09} incurs a proof:code ratio of 20:1 despite the use of automated theorem provers integrated with Isabelle/HOL, e.g., 
Sledgehammer~\cite{sledgehammersmt}. Meanwhile, Ironclad in Dafny~\cite{ironclad14osdi} reports a ratio of 5:1, and EverCrypt in \fstar~\cite{evercrypt} reports a ratio of 3:1, through the use of SMT-solvers and programs that mix computational content and proofs. Improved automation through the use of AI in both settings could help lower these overheads.

In the context of interactive theorem provers, machine-learning has been used to improve premise selection in theorem provers~\cite{kuehlwein2013mash,mikula2023magnushammer}.
Predictive models based on existing proofs to guide proof search have been used in
TacticToe~\cite{gauthier2021tactictoe},
TacTok~\cite{first2020tactok}.
GPT-f~\cite{han2021pact} uses expert iteration
and HTPS~\cite{lample2022hypertree} online reinforcement learning
to improve the model by self-learning based on previous proof attempts.
Baldur~\cite{first2023baldur} uses an LLM-based synthesis and fine-tuned repair model to complete full proofs for Isabelle/HOL theorems.
LeanDojo~\cite{yang2024leandojo} and
LEGO-Prover~\cite{xin2023lego}
integrate retrieval augmentation
to find relevant existing theorems and proofs from the library.
Draft-sketch-proof~\cite{jiang2023draftsketchproof}
introduced the use of informal sketches as an intermediate step,
which has also been used in~\cite{welleck2022naturalprover,huang2024mustard}.

In the context of SMT-assisted program verifiers, 
recent approaches leverage LLMs for synthesizing loop invariants ~\cite{kamath2023finding, liu2023towards, chakraborty2023ranking, pei2023can}.  
Closest to our work, Rakib~\etal~\cite{rakib2024towards} leverage LLMs along with retrieval augmentation to generate both functional specification and code with proofs from natural language in Dafny. 
However, the specifications generated may be too weak or incorrect and requires a human to audit them; this makes it subjective and difficult to treat as a benchmark problem set unlike \fstardataset where specifications are drawn from ground truth.
In addition, prior works on automating proof-oriented programming~\cite{rakib2024towards, yao2023leveraging} are evaluated on much smaller benchmarks with less than 200 mostly introductory algorithmic tasks.
In contrast with \fstardataset we provide a benchmark with more than 32K reproducible \fstar programs and proofs that form part of production software deployed in real-world. 

Additionally, many works use LLMs for general-purpose programming tasks~\cite{ahmad2021unified, chakraborty2022natgen, lahiri2023interactive, chen2021evaluating}. Recent approaches~\cite{endres2023formalizing, key2022speak} even attempt at formalizing functional specifications for languages such as python---without access to static verifiers, these functional specifications often are used as runtime assertions.


\section{Conclusion}
\label{sec:conclusion}

Aiming to enhance both the trustworthiness of AI-generated code and to ease program proof, we investigate AI-based program and proof synthesis backed by a program verifier. While prior work has predominantly focused on tactic-based proof, we investigate AI automation for \fstar, a proof-oriented language with SMT-based automation. To fuel future research in this direction, we introduce the \fstardataset, a public benchmark dataset of \fstar programs and proofs. We expect \fstardataset to evolve and grow: indeed in recent weeks, after our experiments, it has grown to include four more projects, reaching 940K lines of \fstar code and proofs.

On a type-based program and proof synthesis task, we evaluate several state-of-the-art LLMs such as GPT-3.5 and GPT-4, fine-tune smaller language models such as StarCoder. Our findings suggest that LLMs, when trained on appropriate datasets and prompted effectively with necessary information through retrieval, can automate a significant portion programs and proof. However, beyond boosting synthesis performance, many interesting problems remain, notably around specification formulation and modular decomposition of proofs, exciting areas for further exploration.

Overall, with this paper, we contribute to the ongoing discourse on trustworthy AI programming and lay a foundation for advancements in AI-assisted proof-oriented programming.

\bibliographystyle{IEEEtran}
\bibliography{main}

\appendix

\subsection{\fstardataset v2}
\label{v2_data}

Since the initial version of this paper, we have extended our dataset to a larger \fstardataset version 2. As mentioned previously, we envision \fstardataset to be a live, evolving dataset drawn from active \fstar projects. As such, while version 2 primarily includes data from four additional \fstar projects not included in the original version, it also contains data from the original 8 projects, both in the form of new definitions that were previously not present, as well as changes to some of the original definitions. The four new projects are as follows:

\begin{itemize}
\item{Starmada:} a framework for doing proofs by stepwise refinement for concurrent programs in a weak memory model. Starmada is an experimental version of Armada~\cite{lorch20armada}  implemented in \fstar, relying on various advanced features of \fstar's dependent type system for more generic and abstract proofs.

\item{Zeta:} a high performance, concurrent monitor for stateful services proven correct in \fstar and its Steel concurrent separation logic~\cite{arasu23zeta}.

\item{Dice-star:} a verified implementation of the DICE measured boot protocol for embedded devices~\cite{tao21dice}.

\item{Noise-star:} a verified compiler for implementations of Noise protocols, a family of key-exchange protocols~\cite{ho22noisestar}.
\end{itemize}

In total, \fstardataset v2 contains 54,404 definitions drawn from 12 projects, representing around 940,000 lines of \fstar code and proofs. To avoid cross-contamination, we made sure that the F* files that belonged to training dataset in V1 also remained in the training set on V2. The V2 dataset is available in {\tt \url{https://huggingface.co/datasets/microsoft/FStarDataSet-V2}}.


\begin{table}[h]
    \centering
    \caption{\bf Summary statistics of the \fstardataset v2.}
    \label{tab:data_stat_v2}
    \resizebox{\linewidth}{!}{
    \begin{tabular}{lrrrr}
    \hlineB{2}
     \multirow{2}{*}{\bf Metric} & \multirow{2}{*}{\bf Train} & \multirow{2}{*}{\bf Valid} & \multicolumn{2}{c}{\bf Test}\bigstrut[t]\\
     & & & {\bf Intra-project} & {\bf Cross-project} \bigstrut[b]\\
    \hlineB{2}
    Number of Definitions & 30910 & 1955 & 5097 & 16442 \bigstrut[t]\\
    Number of Projects & 6 & 4 & 6 & 2 \\
    Number of Files & 1669 & 125 & 328 & 1179 \bigstrut[b]\\
    \hline
    Avg. num of lines & 4.14 & 3.77 & 3.99 & 3.93 \bigstrut[t]\\
    Avg. num of tokens & 45.34 & 44.90 & 45.56 & 30.30 \bigstrut[b]\\
    \hline
    \# Simply Typed  & 10072 & 834 & 1536 & 7282 \bigstrut[t]\\
    \# Dependently Typed & 15091 & 778 & 2363 & 6067 \\
    \# Proofs & 5747 & 343 & 1198 & 3093 \bigstrut[b]\\
    \hlineB{2}
    \end{tabular}
    }
\end{table}

Table~\ref{tab:data_stat_v2} reports aggregate statistics for V2 dataset. 
\subsection{Synthesis Results on \fstardataset v2}
\label{v2_result}

\para{Improving the RAG} In \S\ref{sec:prompt_v_finetune} we reported that all models performed better on the intra-project examples as compared to the cross-project examples. We indicated that a cause of this might be the manner in which related examples are retrieved, quoting:

\begin{quote}
Both models also perform better on intra-project examples than on cross-project examples. When constructing prompts, we only retrieve related examples from the training data split, which specifically does not include any examples from the cross-project set---as such, the cross-project examples benefit less from retrieval augmentation.
\end{quote}

We have revised the manner in which related examples are retrieved. Rather than restricting related examples to only be fetched from the training data, for a given synthesis problem, we search for examples from all files \emph{that do not depend on} the file of the current problem---this dependence information is available in metadata present in the dataset. We employ the same embedding-based retrieval augmentation technique described in \S\ref{sec:exp}, ranking examples according to their similarity to the goal type and include as many related examples as can fit in the token window, in descending order of similarity score.

\para{Results on V2 with Improved RAG} Table~\ref{tab:results_v2} presents an overview of our results on the \fstardataset-V2 from the fine-tuned model trained on the V1 training set. In addition, \Cref{fig:class_acc_v2} shows performance of different models across different types of problems. We have yet to run all the configurations reported in Table~\ref{tab:prompting_v_finetuning} on the full v2 dataset. We do not report results of the non-fine-tuned versions of Phi-2, Orca-2, and StarCoder on v2, as the performance of these models without fine-tuning on the v1 dataset was not competitive. More notably, we have not yet completed a full run of GPT-4 on the v2 dataset, as a full run is very resource intensive.

\begin{table}[t]
    \centering
    \caption{\bf Synthesis results for \fstardataset v2 }
    \label{tab:results_v2}
    \resizebox{\linewidth}{!}{%
\begin{tabular}{l|ccc|ccc}
\hlineB{2}
 & \multicolumn{3}{c|}{\bf intra-projects (verify@k)} & \multicolumn{3}{c}{\bf cross-project (V@)} \bigstrut[t]\\
\multirow{-2}{*}{\bf Model} & \textbf{k = 1} & \textbf{k = 5} & \textbf{k = 10}& \textbf{k = 1} & \textbf{k = 5} & \textbf{k = 10} \bigstrut[b]\\
\hlineB{2}
GPT-3.5   & 18.85 & 32.33 & 36.59 & 20.53 & 35.87 & 41.63  \bigstrut[t]\\
GPT-4   & xx.xx & xx.xx & xx.xx & xx.xx & xx.xx & xx.xx  \bigstrut[b]\\
\hline
Phi-2     & 24.13 & 35.06 & 39.16 & 24.33 & 36.22 & 40.59  \bigstrut[t]\\
Orca-2    & 18.91 & 31.76 & 37.08 & 15.89 & 29.07 & 34.42  \\
StarCoder & 36.85 & 50.93 & 56.07 & 38.58 & 53.51 & 58.13  \bigstrut[b]\\
\hlineB{2}
\end{tabular}%
}
\end{table}


\begin{figure}[tbh]
    \centering
    \begin{subfigure}[t]{0.90\linewidth}
        \centering
        \includegraphics[width=0.95\linewidth]{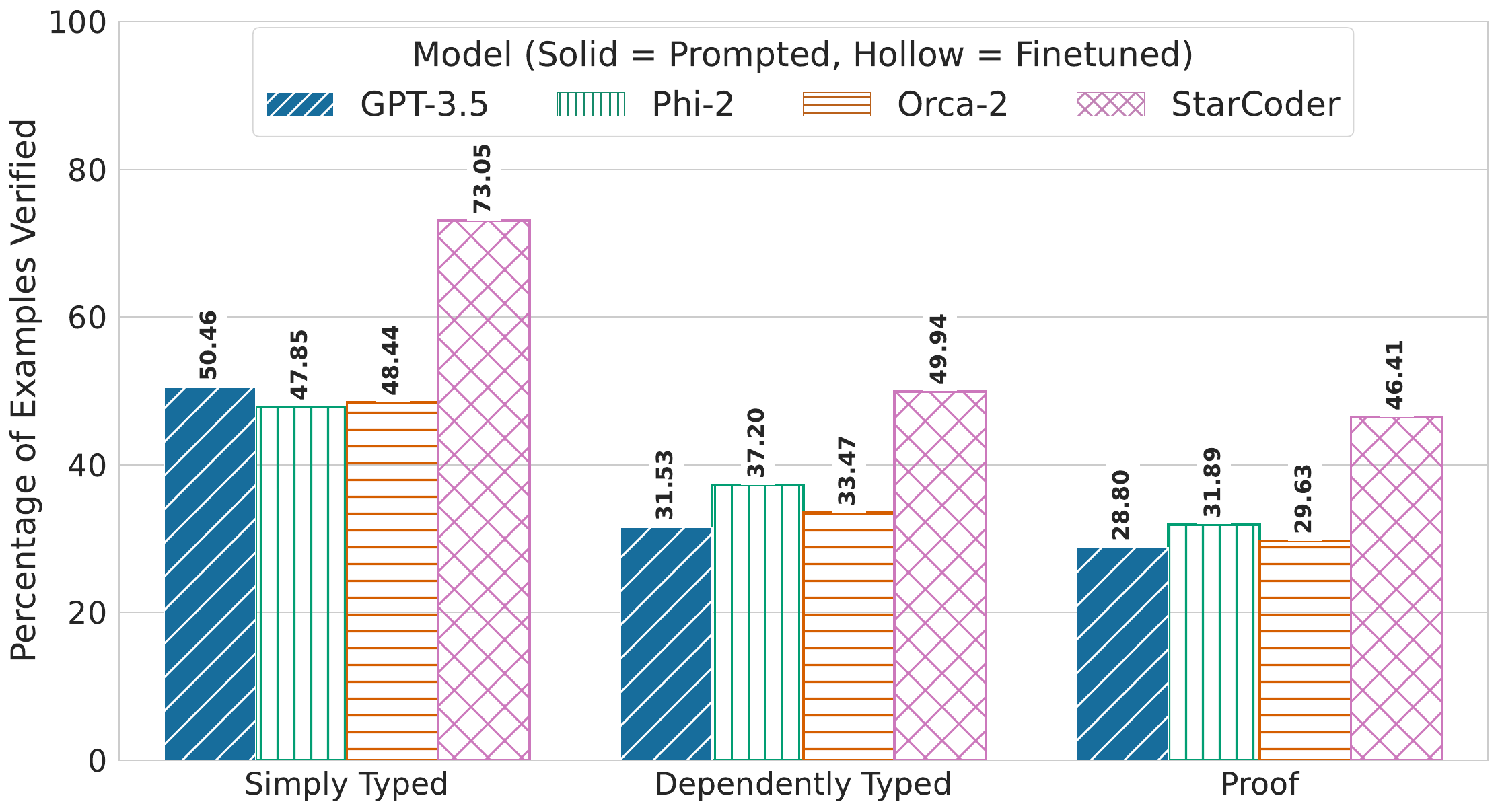}
        \caption{Intra Project evaluation}
        \label{subfig:rq1_class_test_v2}
    \end{subfigure}
    \begin{subfigure}[t]{0.90\linewidth}
        \centering
        \includegraphics[width=0.95\linewidth]{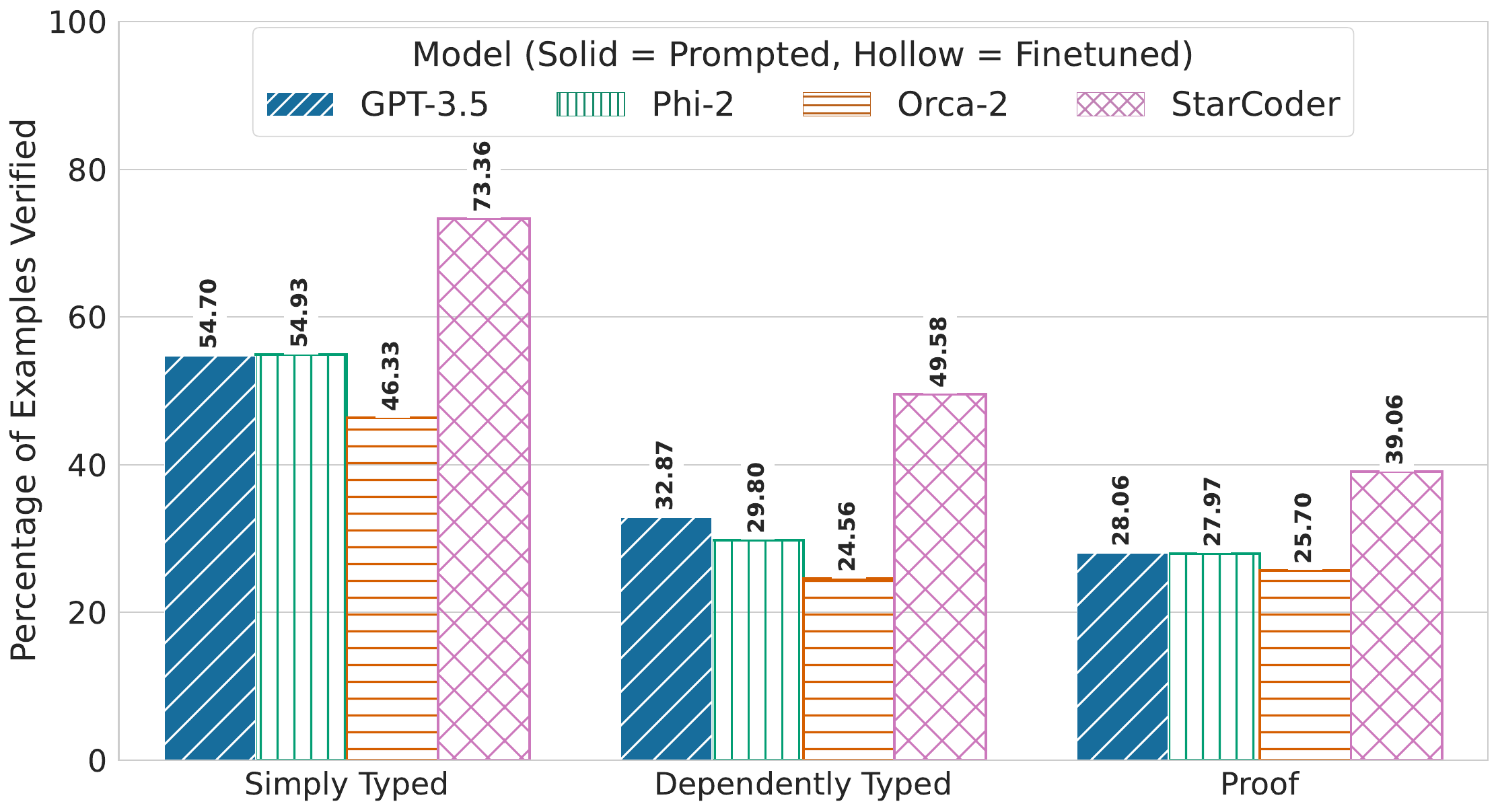}
        \caption{Cross Project evaluation}
        \label{subfig:rq1_class_challenging_v2}
    \end{subfigure}
    \caption{\bf Verify@10 across different types of examples for different models in the V2 dataset.}
    \label{fig:class_acc_v2}
\end{figure}

Our results make it clear that fine-tuned models are not over-fitted to a particular version of the dataset. Our prior results extend well to the new dataset, and indeed with our new retrieval augmentation strategy, the difference in performance between the intra-project and cross-project classes is not significant. 
Perhaps surprisingly, the performance across the board on the cross-project class is better than in the intra-project class: this is explained by noting that the cross-project class contains a larger proportion of simply typed definitions (44.3\%) that the intra-project class (33.5\%). 
In addition, cross-project examples are shorter than that of Intra-Project examples.

\end{document}